\documentclass[letterpaper,twocolumn,10pt]{article}
\usepackage{usenix-2020-09}

\usepackage{tikz}
\usetikzlibrary{patterns}
\usepackage{amsmath}
\usepackage{amsfonts}

\usepackage{filecontents}

\usepackage{algorithm}
\usepackage[noend]{algpseudocode}

\usepackage{pgfplots}
\usepackage{enumitem}
\usepackage{cleveref}
\usepackage{titlesec}

\usepackage{xcolor}

\usepackage{changepage} 

\usepackage{comment}
\usepackage{soul}
\usepackage{dirtytalk}

\usepackage{caption}
\usepackage{subcaption}
\usepackage{float}

\usepackage{pf2}

\makeatletter
\renewcommand\paragraph{\@startsection{paragraph}{4}{\z@} {2pt} {-12pt} {\normalfont\normalsize\bfseries}} \makeatother
\titlespacing\section{0pt}{6pt plus 2pt minus 2pt}{3pt plus 1pt minus 1pt}
\titlespacing\subsection{0pt}{4pt plus 1pt minus 1pt}{2pt plus 1pt minus 1pt}
\titlespacing\subsubsection{0pt}{1pt plus 0pt minus 0pt}{1pt plus 0pt minus 0pt}

\pflongnumbers

\newcommand{\ProtocolX}{Alea-BFT} 

\newcommand{\fillgap}[2]{$\langle \texttt{FILL-GAP}, #1, #2 \rangle$}
\newcommand{\filler}[1]{$\langle \texttt{FILLER}, #1 \rangle$}

\newif\ifwip
\wipfalse 

\ifwip
\newcommand{\rrnote}[1]{\textbf{[RR: #1]}}
\newcommand{\aonote}[1]{\textbf{[AO: #1]}}
\newcommand{\abnote}[1]{\textbf{[AB: #1]}}
\newcommand{\dsanote}[1]{\textbf{[DSA: #1]}}
\newcommand{\mfnote}[1]{\textbf{[MF: #1]}}

\newcommand{\review}[1]{#1}
\newcommand{\reviewdel}[1]{}

\newcommand{\freview}[1]{\textcolor{blue}{\textbf{#1}}}
\newcommand{\freviewdel}[1]{\textcolor{red}{\st{#1}}}

\else

\newcommand{\rrnote}[1]{}
\newcommand{\aonote}[1]{}
\newcommand{\abnote}[1]{}
\newcommand{\dsanote}[1]{}
\newcommand{\mfnote}[1]{}

\newcommand{\review}[1]{#1}
\newcommand{\reviewdel}[1]{}
\newcommand{\freview}[1]{#1}
\newcommand{\freviewdel}[1]{}

\fi

\soulregister{\ProtocolX}{7}
\soulregister{\Cref}{7}
\soulregister{\cite}{7}

\begin{document}

\date{}

\title{\Large \bf \ProtocolX: Practical Asynchronous Byzantine Fault Tolerance
}

\author{Diogo S. Antunes,
Afonso N. Oliveira,
André Breda,\\
Matheus Guilherme Franco,
Henrique Moniz,
Rodrigo Rodrigues\thanks{A.\ Oliveira is now with Three Sigma. M.\ Franco is now with ssv.network.\vspace{.4em}\\This is an extended version of the paper appearing in~\href{https://www.usenix.org/conference/nsdi24}{NSDI'24.}}\\
Instituto Superior T\'{e}cnico (ULisboa) and INESC-ID
} 

\maketitle

\begin{abstract}

Traditional Byzantine Fault Tolerance (BFT) state machine replication protocols assume a partial synchrony model, leading to a design where a leader replica drives the protocol and is replaced after a timeout. Recently, we witnessed a surge of asynchronous BFT protocols, which use randomization to remove the need for bounds on message delivery times, making them more resilient to adverse network conditions. However, existing research proposals still fall short of gaining practical adoption, plausibly because they are not able to combine good performance with a simple design that can be readily understood and adopted. In this paper, we present \ProtocolX, a simple and highly efficient asynchronous BFT protocol, which is gaining practical adoption, namely in Ethereum distributed validators. \ProtocolX\ brings the key design insight from classical protocols of concentrating part of the work on a single designated replica\freviewdel{,} and incorporates this principle in a simple two\freviewdel{ }\freview{-}stage pipelined design, with an efficient broadcast led by the designated replica, followed by an inexpensive binary agreement.
\freviewdel{We evaluated our research prototype implementation and two real-world integrations in cryptocurrency ecosystems, and our results show excellent performance, improving on the fastest protocol (Dumbo-NG) in terms of latency, and also good performance under faults.}
\freview{The evaluation of our research prototype implementation and two real-world integrations in cryptocurrency ecosystems shows excellent performance, improving on the fastest protocol (Dumbo-NG) in terms of latency and displaying good performance under faults.}
\end{abstract}

\section{Introduction}\label{sec:intro}

The history of Byzantine fault tolerant (BFT) replication has gone through different stages throughout the years, from the initial exploration of the topic in the 1980s~\cite{Lamport}\freviewdel{,} to the start of a series of practical protocols that achieve good performance in the late 1990s~\cite{pbft}, and more recently the real-world adoption of this class of protocols in the context of cryptocurrencies and blockchains~\cite{hotstuff}.

BFT protocols must carefully navigate the constraints of the FLP impossibility result~\cite{flp}. This result states that no deterministic algorithm can guarantee consensus (or, equivalently, agreement on the outcome of a client request within a replicated state machine) in a fully asynchronous system where even a single process might experience a crash failure.
For many decades, the almost universally accepted way to circumvent this hurdle was by assuming a partial synchrony model, where the network is assumed to be initially asynchronous but, after an unknown point in time, delivers and processes messages within a certain time bound~\cite{partialsynch}. This\freview{ model} leads to a class of protocol designs where a leader can drive the execution of the protocol. In this case, after a timeout \freviewdel{that indicates}\freview{indicating} that the protocol is not making progress, all replicas must cooperate in picking a new leader.

Recently, researchers picked up a different line of research that had been somewhat dormant for many years: asynchronous BFT protocols~\cite{aspnes2003randomized}. These protocols are safe and live irrespectively of any timing assumptions being met, but at the cost of probabilistic guarantees, i.e., they are provided with very high probability. Removing these timing assumptions \freviewdel{brings the advantage of allowing the protocol to be more resilient}\freview{improves protocol resilience} against \reviewdel{node}\review{replica} and network delays, which may be due to reasons ranging from network problems to malicious activity~\cite{singh2008bft,clement2009aardvark}.
The recent surge of interest in asynchronous BFT came after the publication of a protocol called HoneyBadgerBFT (HBBFT)~\cite{honeybadger}.
\reviewdel{, which is usually regarded as the first practical asynchronous BFT protocol.} Since \review{its} publication, several other protocols appeared~\cite{speeding-dumbo,dispersed-ledger,allyouneedisdag,bolt-dumbo,j-ditto,dumbo,beat,epic}, making tremendous progress in the properties of these protocols, namely their performance and asymptotic complexity.

However, while these proposals succeeded in showing that asynchronous BFT algorithms can perform well, they \freviewdel{are}\freview{have} yet to gain practical adoption in production systems. In our view, this can largely be due to the fact that existing protocols fall short of striking a virtuous combination of good performance and simple protocol design. The academic research community often overlooks the latter, but it can be a decisive factor in practical adoption. An illustrative example of this point, from the partially synchronous arena, is the work of Istanbul BFT~\cite{moniz2020ibft} (also known as QBFT~\cite{qbft-web}). This protocol is widely adopted by the blockchain community~\cite{qbft-adoption}, to a large extent due to it being simple to understand and implement, and despite it being published more than two decades after PBFT~\cite{pbft} and its many successors.


In this paper, we present \ProtocolX, the first protocol for asynchronous BFT state machine replication that brings together top-notch performance~--~in terms of throughput, latency, and asymptotic complexity~--~with a simple and elegant design and practical adoption in real-world systems.

The main insight in \ProtocolX\ is that it selectively brings a key design feature from classical partially synchronous protocols, namely having a per-request designated leader replica that drives the protocol execution for that request.
\review{To avoid resorting to timeouts for leader replacement,} the choice of leader can constantly rotate among all replicas, as previously done in the crash~\cite{mao2008mencius} and Byzantine~\cite{veronese2009spin} models.
Then, by splitting the request execution into two phases\freviewdel{,} and placing on this replica the responsibility of initiating the broadcast phase to disseminate client requests, \ProtocolX\ avoids redundant instances of expensive building blocks \review{present in existing} asynchronous protocols and also avoids the use of threshold cryptography to encrypt proposals replicated across processes.
However, this also introduces challenges, namely that there is no guarantee that the broadcast by the leader will reach a sufficient number of replicas in time for the subsequent agreement phase.
We address this challenge by including an agreement phase, pipelined with the broadcast phase, whose goal is to allow replicas to agree on whether it is safe to execute the client request. The execution can proceed if sufficient replicas received the request to reconstruct it. Otherwise, the request is locally stored in one of the queues of pending requests.
\review{This leads to a design featuring a novel combination of existing building blocks, namely using VCBC as a broadcast primitive and ABA as the driver for agreement, which are judiciously joined together to provide a simple and performant protocol.}

    



We report on three implementations of \ProtocolX: a research prototype and two real-world implementations, one of them in the context of \review{the SSV} Ethereum distributed validator (the key technology behind staking pools), which is currently being considered to replace QBFT as its main consensus protocol in the near future~\cite{ssv-blog}, and another in the context of \reviewdel{the future}\review{an experimental} consensus layer for the subnets of \review{Filecoin~\cite{mir_readme}}.
More recently, a second Ethereum distributed validator incorporated \ProtocolX\ in its protocol roadmap~\cite{obol-roadmap}. 


Our experimental evaluation of these three prototypes shows that \ProtocolX\ has excellent performance,  \review{namely with comparable throughput and} better latency than the fastest available asynchronous BFT from the recent literature~\cite{dumbo-ng}.
This combination of excellent performance, protocol elegance, and real-world adoption makes \ProtocolX\ a practical solution for asynchronous BFT.

The remainder of the paper is organized as follows. \Cref{sec:related-work}  surveys related work. \Cref{sec:basics} describes the system model and building blocks.
\Cref{sec:protocol} presents the design of \ProtocolX, and we optimize it in \Cref{sec:opt}. \Cref{sec:efficiency} analyses its asymptotic complexity. \Cref{sec:corr} sketches a correctness proof. \Cref{sec:impl} describes our various implementations, which are  evaluated in \Cref{sec:eval}. We conclude in \Cref{sec:conclusion}.

\section{Related Work}\label{sec:related-work}

The Byzantine consensus problem was formulated by Lamport et al.~\cite{Lamport}, and\freview{,} over time\freview{,} \freviewdel{led}\freview{accumulated} a large body of research in the area~\cite{rampart,securering,BQS,pbft,hq,zyzzyva}.
BFT recently gained adoption in cryptocurrencies and blockchains, with several new protocols for those deployments~\cite{hotstuff,moniz2020ibft}.

From these, the protocols that implement a form of consensus~--~namely state machine replication protocols~\cite{schneider1990implementing}~--~face the FLP impossibility of consensus in asynchronous systems~\cite{flp}.
To circumvent this result, most BFT systems rely on timing assumptions such as partial synchrony~\cite{partialsynch} for liveness. This is the case, for instance, of systems such as PBFT~\cite{pbft} and also more recent proposals such as HotStuff~\cite{hotstuff}, Kauri~\cite{kauri} or ISS~\cite{iss}. Partially synchronous protocols can, however, be sensitive to conditions like a primary that deliberately slows down the system~\cite{clement2009aardvark} or situations where replicas are correct but the network is unreliable~\cite{singh2008bft}. 


As an alternative to assuming partial synchrony, randomized protocols circumvent FLP by guaranteeing the liveness property with high probability. The design for this class of protocols runs the main algorithm through multiple rounds until its nondeterministic nature allows the probability of not having liveness to be irrelevant.
These protocols can then operate over a fully asynchronous model, \freviewdel{therefore }eliminating the need for timing assumptions.

Existing asynchronous BFT protocols do not simultaneously achieve the goals of simplicity and performance, which are key for practicality. In particular, the initial asynchronous BFT protocols~\cite{rabin1983randomized,bracha1987asynch,ben1994asynchronous,cachin2001secure,moniz2008ritas} are very elegant (sometimes described in less than 10 lines of pseudocode~\cite{bracha1987asynch})\freviewdel{,} but have high communication costs and expected termination time.
More recently, several new randomized protocols appeared. At the core of this new line of proposals is an asynchronous binary agreement (ABA) primitive, in which processes decide on the value of a single bit.
These ABA protocols are then used as building blocks for atomic broadcast and state machine replication solutions. After a small set of initial proposals, namely HoneyBadgerBFT (HBBFT)~\cite{honeybadger}, BEAT~\cite{beat}, EPIC~\cite{epic}, and Dumbo~\cite{dumbo}, a large number of proposals emerged over the last few years~\cite{speeding-dumbo,dispersed-ledger,allyouneedisdag,bolt-dumbo,j-ditto}. Given the relatively large literature, we only describe in detail two of these proposals, namely the pioneering work of HBBFT~\cite{honeybadger} and a recent proposal \review{with excellent performance} named Dumbo-NG~\cite{dumbo-ng}.
\reviewdel{
, which, to our knowledge, achieved the best performance as of today.}


HBBFT~\cite{honeybadger} is based on the observation that atomic broadcast can be built on top of an asynchronous common subset (ACS) framework by combining it with a threshold encryption scheme. In ACS\freview{,} every party proposes an input value\freviewdel{,} and outputs a common vector containing the inputs of at least $N-f$ distinct parties.
HBBFT constructs ACS from the composition of two phases: reliable broadcast (RBC) and asynchronous binary agreement (ABA). 
During the broadcast phase, every replica starts an RBC instance to disseminate its proposal to all other replicas. Then, in the agreement phase, $N$ parallel ABA instances are invoked to decide on an $N$-bit vector, where the $i$-th value indicates whether or not to include the proposal from replica $P_i$ in the final ACS output. Here, threshold encryption prevents an adversary from selectively censoring requests\freviewdel{,} by selecting which proposals to include in the ACS output vector.

To our knowledge, the best performing and state-of-the-art proposal in this area (outperforming its competitors by several-fold) is Dumbo-NG~\cite{dumbo-ng}. 
This protocol decouples a \review{continuously} running broadcast phase from a sequence of multi-valued Byzantine agreement (MVBA) instances.
The broadcast phase uses a custom protocol that resembles VCBC (see Section~\ref{sec:basics}), whereas the MVBA phase reuses an existing protocol, whose validity predicate is fine-tuned to check for valid threshold signatures and other protocol-specific conditions. The presence of an MVBA protocol introduces an $\mathcal{O}(n^3)$ message complexity, which contrasts with \ProtocolX's use of a round\freviewdel{ }\freview{-}robin ABA, with only $\mathcal{O}(n^2)$ complexity.


Generally, we can categorize previous proposals as either suffering from high communication costs (pre-HBBFT protocols) or having a more complex design that hinders practical adoption (new generation, starting from HBBFT). In contrast, \ProtocolX\ brings together a simple and elegant design with excellent performance\freviewdel{,} and is now being adopted in real-world systems, namely Ethereum distributed validators. \review{This might be in part due to the simplicity of the protocol and its components -- for instance, while Dumbo-NG uses an MVBA, which is complex in both the provided guarantees and its implementation, \ProtocolX\ leverages a much simpler ABA primitive, resulting in an overall protocol that is easier to understand and implement}. Furthermore,
\ProtocolX\ improves on most prior asynchronous protocols through its near quadratic message complexity. Note that while quadratic protocols have been \review{theoretically} proposed~\cite{quadraticBFT}, \review{we do not know of any protocol with such characteristics that was} implemented.



\section{Basics}\label{sec:basics}
In this section, we present the  system model and precisely define the basic blocks upon which \ProtocolX\ is built.

\subsection{System model}
We consider a distributed system composed of $N$ processes, also called replicas, uniquely identified from the set $S = \{P_0, ..., P_{N-1}\}$ and an arbitrary number of clients.

We assume a Byzantine failure model where up to $f=\lfloor\frac{N-1}{3}\rfloor$ replica processes can fail arbitrarily during the execution of the protocol. The remaining processes follow the protocol specification and are termed correct. \ProtocolX\ is adaptively secure against an adversary that dynamically determines the replicas to compromise. That said, it reuses two classes of protocols described later in this section, which can have either statically secure or adaptively secure instantiations. As such, choosing a statically secure subprotocol would downgrade the solution to be statically secure. 

The system is asynchronous, with the message delivery schedule under adversarial control, and without bounds on communication delays or processing times. Processes are fully connected by channels, providing guarantees that messages are not modified in transit and are eventually delivered. In practice, this requires message retransmission and point-to-point authentication, but by considering this network model, we can omit these from the protocol description.

Lastly, the adversary is assumed to be computationally bound and thus unable to subvert cryptographic primitives.

\subsection{Specification}

We \review{specify}\reviewdel{developed} \ProtocolX\ as an atomic broadcast protocol, which is a \review{common} abstraction for implementing state machine replication. Intuitively, this allows a process (e.g., a proxy replica) to broadcast a message (e.g., a client request\footnote{Client requests are also referred to in the literature as state machine commands. Throughout the paper, we will use only the term \emph{request}.} to be executed on the state machine) to all processes, ensuring that all processes deliver all messages in the same order (executing all client requests in the same order and therefore transitioning through the same sequence of states). Formally, atomic broadcast is defined as follows (with the standard assumption that messages include a per-sender id and sequence number to make them unique)~\cite{Hadzilacos1994AMA}:
\begin{itemize}[leftmargin=*,noitemsep,nolistsep]
    \item \textit{Validity.}
    If a correct process broadcasts a message $m$, then some correct process eventually delivers $m$.

    \item \textit{Agreement.}
    If any correct process delivers a message $m$, then every correct process delivers $m$.

    \item \textit{Integrity.}
    A message $m$ appears at most once in the delivery sequence of any correct process.

    \item \textit{Total order.}
    If two correct processes deliver messages $m$ and $m'$, then both deliver $m$ and $m'$ in the same order.
\end{itemize}


\subsection{Building blocks}
\ProtocolX\ is designed in a modular way by reusing several subprotocols to carry out certain tasks.
In this modular architecture, upper-level protocols provide inputs and receive outputs from subprotocols at the lower layers.
Next, we present the precise specification of these underlying primitives.

\subsubsection{Verifiable Consistent Broadcast Protocol}
\reviewdel{protocol to deliver a payload message from a distinguished sender to all replicas.}
Verifiable consistent broadcast (VCBC) is a \review{broadcast variant that was first proposed by Cachin et al.~\cite{cachin2001secure}}. It can only guarantee that all correct replica processes deliver the broadcast value if the sender is correct; however, it always ensures that no two correct processes deliver conflicting messages.
Additionally, it allows any party $P_i$ that has delivered \reviewdel{the payload} message $m$ to inform another party $P_j$ about the outcome of the broadcast execution, allowing it to deliver $m$ immediately and terminate the corresponding VCBC instance.
More formally, a VCBC protocol ensures the following properties~\cite{cachin2001secure}:
\begin{itemize}[leftmargin=*,noitemsep,nolistsep]
    \item \textbf{Validity:}
    If a correct sender broadcasts $m$, then all correct parties eventually deliver $m$.
    
    \item \textbf{Consistency:}
    If a correct party delivers $m$ and another \review{correct} party delivers $m'$, then $m = m'$.
    
    \item \textbf{Integrity:}
    Every correct party delivers at most one message. Additionally, if the sender is correct, then it previously broadcast the message.

    \item \textbf{Verifiability:}
    If a correct party delivers a message $m$, then it can produce a single protocol message $M$ that it may send to other parties such that any correct party that receives $M$ can safely deliver $m$.
    
    \item \textbf{Succinctness:}
    The size of the proof $\sigma$ carried by $M$ is independent of the length of $m$.
\end{itemize}
In \ProtocolX\freview{,}\ we use a VCBC implementation consisting of extending an echo broadcast protocol~\cite{cachin2001secure} with threshold signatures to generate \reviewdel{or validate}the proof $\sigma$. \reviewdel{associated with $M$.}In short, the protocol consists of the distinguished sender process sending $m$ to all processes and collecting a Byzantine quorum of $\lceil\frac{n+f+1}{2}\rceil$ signature shares in the replies, allowing the sender to combine these shares and convey the signature to all processes in the final message step.
Using threshold signatures keeps the message size constant, ensuring succinctness. The message complexity of the VCBC protocol we use is $\mathcal{O}(N)$ and its communication \freview{complexity }is $\mathcal{O}(N(|m| + \lambda))$, assuming the size of a threshold signature and share is at most $\lambda$ bits.

\subsubsection{Asynchronous Binary Agreement}
An asynchronous binary agreement (ABA) protocol allows correct processes to agree on the value of a single bit. Each process $P_i$ proposes a binary value $b_i \in \{0, 1\}$ and decides\freviewdel{ for} a common value $b$ from the set of proposals by correct processes. Formally, a binary agreement protocol can be defined by the following properties:
\begin{itemize}[leftmargin=*,noitemsep,nolistsep]
    \item \textbf{Agreement:}
    If any correct process decides $b$ and another correct process decides $b'$, then $b=b'$.
 
    \item \textbf{Termination:}
    Every correct process eventually decides.

    \item \textbf{Validity:}
    If all correct processes propose $b$, then any correct process that decides must decide $b$.
\end{itemize}
Given the FLP  theorem~\cite{flp}, no deterministic algorithm can satisfy all the previous properties in the asynchronous model of \ProtocolX. As such, we use a randomized solution with the following termination property:
\begin{itemize}[leftmargin=*,noitemsep,nolistsep]
    \item \textbf{Termination:}
    The probability that a correct process is undecided after $r$ rounds approaches $0$ as $r$ approaches $\infty$.
\end{itemize}
This way, even though the number of rounds required to reach agreement is unbounded, the probability that the protocol does not terminate converges to zero.

We instantiate this primitive via the Cobalt ABA~\cite{macbrough2018cobalt} protocol, a modified version of the proposal by Most\'{e}faoui et al.~\cite{mostefaoui2014signature}.
The protocol relies on a common source of randomness, i.e., a “common coin”, realized from a threshold signature scheme by signing a unique bit string corresponding to the name of the coin and combining the signature shares to generate a random seed~\cite{cachin2005random}.
The protocol proceeds in rounds, each consisting of the following all-to-all message communication steps: \textsc{init}, conveying the most recent proposal (0 or 1) of each process,
followed by \textsc{aux} and \textsc{conf}, trying to confirm the existence of strong support (i.e., a Byzantine quorum) in the previous step for a value. At the end of these exchanges, processes either decide a value (if that support was gathered) and convey it through a \textsc{finish} message or otherwise move to the next round, changing their proposal to the value of the common coin whenever both final outcomes are considered possible.
This protocol provides optimal resilience, $\mathcal{O}(N^2)$ expected message complexity, $\mathcal{O}(\lambda N^2)$ expected communication complexity and terminates in $\mathcal{O}(1)$ expected time.
\section{\ProtocolX}\label{sec:protocol}

This section presents \ProtocolX, starting with a broad overview, followed by a detailed description and pseudocode. 

\subsection{Overview}\label{sec:protocol:overview}
One of the central insights of  \ProtocolX\ is to have a single replica propose a value per consensus instance, similar to what happens in leader-based protocols in the partially synchronous model, \freviewdel{while}\freview{and} all others agree on whether to deliver it or not. Departing from a design where all replicas try to insert each client request in the total order enables us to remove an all-to-all communication phase and only have a single ABA execution per client request (or batch of requests). This\freview{ insight} then leads to the following initial design.

\noindent \textbf{Strawman Proposal.}
The first design consists of adapting the ACS construction of HBBFT\freviewdel{,} but\freview{,} instead of having all replicas simultaneously propose values, a single replica is selected as the proposer for each consensus round. The role of the proposer is to \freviewdel{select}\freview{choose a value (or batch of values)} from its buffer of pending requests\reviewdel{ a value (or batch of values)} to serve as a proposal and broadcast it to all replicas, using a broadcast primitive that ensures that all replicas receive the same value~--~if they output a value at all~--~a property ensured by consistent broadcast~\cite{cgr:book}.
Correct replicas would then proceed to execute a \textit{single} ABA to determine whether to deliver the proposed value for that round (if enough replicas have received it to ensure it persists despite faults or asynchrony) or not deliver anything.
Additionally, the proposer is deterministically rotated upon every ABA execution \freviewdel{in order }to address the scenario where the proposer is faulty without introducing a fail-over sub-protocol, similar to what happens in other protocols for the partially synchronous model (both with crash~\cite{mao2008mencius} and Byzantine~\cite{veronese2009spin} faults) that incorporate leader rotation into the normal operation, such that it is constantly changing.

This strawman protocol, however, raises an immediate problem. In previous protocols based on an ACS framework, replicas are guaranteed to receive proposals from at least $N-f$ correct replicas. Therefore\freview{,} they can wait until this threshold is met before deciding which values to input for the subsequent agreement stage. In contrast, in our strawman protocol, only a single replica takes the role of the proposer at any given time, so there is no way to determine whether the current proposer is faulty or not, thus making it difficult to decide which value to input into the ABA without resorting to some timeout, which contradicts the asynchronous model.

\noindent \textbf{Final design.}
The impossibility of waiting for some threshold to be met before deciding the value to input to the ABA stage leads us to the insight of not waiting at all and instead allowing undelivered proposals to exist, which are then carried \review{over} across rounds. In other words, every time a particular replica is reelected as the proposer, the corresponding ABA execution will decide over its queue of pending proposals instead of a single newly proposed value (or batch of values).
This way, replicas can submit their input to start the ABA for a new round as soon as they conclude the previous round, since even if the decision is 0 (i.e., not deliver any proposal in the round), the same proposal will be eventually revisited when the same replica becomes the leader and a larger threshold of replicas become aware of the proposal, guaranteeing convergence to an ABA decision of 1 over time. The ABA execution also serves \freview{as }a synchronization mechanism between replicas\freviewdel{,} since no replica can progress to a round until it \freviewdel{has concluded}\freview{concludes} all ABA instances for previous rounds.

In \ProtocolX, we leverage this idea to decompose the monolithic architecture of previous ACS-based protocols, in which a binary agreement instance actively waits for the corresponding broadcast to terminate, into a two\freviewdel{ }\freview{-}stage pipeline, where the results of the first phase (broadcast component) are queued to be eventually processed, either by the current or by a subsequent execution of the second phase (agreement component). Very importantly for performance, these two phases are executed in parallel, allowing for efficient pipelining.
\begin{figure}[t]
\begin{center}
\includegraphics[width=.8\linewidth]{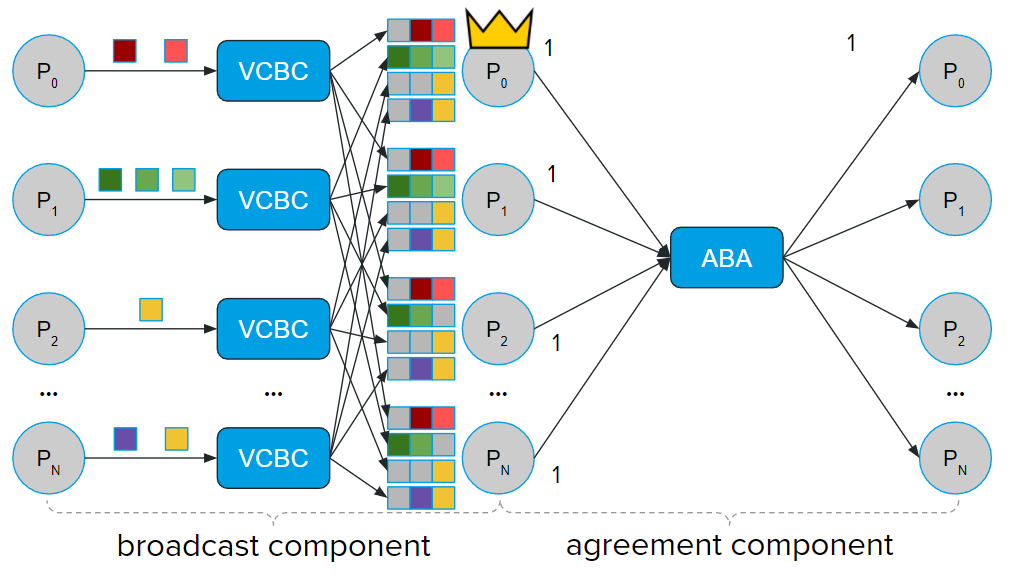}\vspace{-1em}
\end{center}
\caption{Overview of \ProtocolX. Requests go through a single broadcast primitive (VCBC), are inserted in a priority queue at each replica, determining the final ABA input.\vspace{-1em}}
\label{fig:alea}
\end{figure}

Figure~\ref{fig:alea} depicts the resulting overall protocol flow. It starts with the broadcast component of the \ProtocolX\ pipeline, where replicas receive client requests, (optionally) batch them by storing these in a pending buffer of size $B$, and, when the buffer is full, disseminate its contents via a VCBC primitive tagged with an incremental sequence number $s$. The output of VCBC at each replica is stored in a buffer and only removed upon a decision of 1 in the subsequent phase. 
\review{The broadcast} stage produces an instance of an ordered \reviewdel{backlog}\review{queue} of undelivered proposals at each replica. 
These instances are then used as input to the next component of the pipeline. Note that every replica maintains $N$ \review{queues}\reviewdel{backlog} of undelivered proposals, one for each replica in the system, and these grow and shrink over time depending on how efficiently the agreement component can process them.

The next stage is the agreement component, which iteratively selects one of the queues and decides whether to deliver the oldest proposal.
To \freviewdel{do so}\freview{this end}, replicas participate in a single ABA execution, voting $1$ if their queue contains this proposal or $0$ otherwise.
If the decision is $1$, \freviewdel{this indicates that}\freview{then} a sufficient threshold of correct replicas are aware of the proposal and may safely deliver it, as \review{explained next.}\reviewdel{all other replicas are guaranteed to be able to fetch it if needed, via a recovery mechanism actively.}
Otherwise, if a decision is $0$, the agreement component simply moves on to the next queue, repeating the same process\freviewdel{ again}.

As mentioned, since the broadcast's VCBC primitive may terminate at different times in different processes, we need to address the scenario where a correct process outputs an ABA decision of 1, but does not yet know the corresponding proposal.
In this scenario, such a correct process requests the missing proposal from the other processes that voted 1. This\freview{ recovery mechanism} is guaranteed to work for the following reasons. Since ABA decided \freviewdel{a value of }1, \reviewdel{this implies that}at least one correct process voted for 1. Therefore, this process \review{has the required VCBC proof (as guaranteed by VCBC's verifiability property) and} can\reviewdel{compute a VCBC proof for the pending proposal and} forward it to the requesting process.

\subsection{Detailed description}
\reviewdel{Next we explain the \ProtocolX\ protocol in detail.}
Processes \review{in \ProtocolX\ } maintain two state variables shared between the two components of the pipeline: variable $S_i$, consisting of the set of all messages delivered by the protocol, which is initialized as empty upon a call to the \textproc{Start} procedure, and updated during the execution of the agreement component; and variable $\texttt{queues}_i$, comprising an array of $N$ priority queues, each corresponding to a distinct replica $P_x, \forall x \in \{0, ..., N-1\}$.
Algorithm~\ref{algo:alea:start} is responsible for initializing the shared state variables and starting the pipeline components upon a call to the \textproc{Start} procedure. In the remainder of this section, we \freviewdel{start}\freview{begin} by specifying the data structure of the priority queues and then describe the two components of the pipeline in turn.

\begin{algorithm}[h]
\begin{algorithmic}[1]
\small

\algdef{SN}[constants]{Constants}{EndConstants}
{\textbf{constants:}}

\algdef{SN}[variables]{Variables}{EndVariables}
{\textbf{state variables:}}

\Statex
\Constants
    \State $N$ 
    \State $f$ 
\EndConstants

\vspace{-7px}

\Statex
\Variables
    \State $S_i \gets \emptyset$ 
    \label{algo:vars:s}
    \State $queues_i \gets \emptyset$ 
    \label{algo:alea:vars:queues}
\EndVariables

\vspace{-7px}

\Statex
\Procedure{START}{}
    \State $queues_i[x] \gets$ \textbf{new} pQueue() $,  \forall x \in \{0, ..., N-1\}$
    \State \textbf{async} \texttt{BC-START}() 
    \State \textbf{async} \texttt{AC-START}() 
\EndProcedure
\end{algorithmic}
\caption{\ProtocolX\ - Initialization (at $P_i$)}
\label{algo:alea:start}
\end{algorithm}

\subsubsection{Priority queues}
A priority queue is a custom data structure for storing elements\freviewdel{,} sorted according to their priority values.
We refer to each position in a priority queue as a slot, uniquely identified by a priority value associated with it, where the lower-numbered priority values represent the elements that must be processed first. Only a single element can be inserted in a given slot, even after being removed, as the slot is permanently label\freview{l}ed as used and cannot store another element.
A special slot called the head slot always points to \review{the slot with the lowest-numbered priority} whose value has not been removed yet. The pointer to the head slot progresses incrementally, conditioned by the \reviewdel{insertion and} removal of elements from the queue.
A priority queue exposes the following attributes:
\begin{itemize}[leftmargin=*,noitemsep,nolistsep]
    \item \textbf{id:}
    The unique identifier of the queue (static).
    
    \item \textbf{head:}
    The priority value associated with the head slot of the queue (dynamic).
\end{itemize}
Additionally, a priority queue provides an interface for accessing and modifying its contents as described below:
\begin{itemize}[leftmargin=*,noitemsep,nolistsep]
    \item \textbf{Enqueue $(v, s)$:}
    Add an element $v$ with a given priority value $s$ to the queue (ignored if the corresponding slot is not empty).
  
    \item \textbf{Dequeue $(v)$:}
    Removes all instances of the specified element $v$ from the queue\freviewdel{,} if it is present.
  
  
    \item \textbf{Peek $() \rightarrow \{v, \bot \}$:}
    Retrieve, without removing, the element $v$ in the head slot of the queue\freviewdel{,} or $\bot$ if the slot is \review{still} empty \review{(because no \textbf{Enqueue} for that slot has been invoked yet)}.
\end{itemize}
As we will see, \ProtocolX\ leverages the properties of this structure to mediate the communication between the broadcast and agreement components of the protocol pipeline. In particular, each of the $N$ priority queues that each replica maintains keeps track of the undelivered proposals \freviewdel{coming}\freview{originating} from the other replicas, ordered by the priority value assigned to those proposals.

\subsubsection{Broadcast Component}
The broadcast component is responsible for establishing an initial local order over the client updates received and propagating that order to other replicas.
Every replica process maintains \freviewdel{a }two local state variables, a buffer of pending client requests $buf_i$, and an integer value $priority_i$, indicating the next sequence number it should assign to a proposal.
The main logic of this component, illustrated in Algorithm~\ref{algo:alea:bc}, is split between two upon rules:

\noindent \textbf{Upon rule 1 (\crefrange{algo:bc:upon:input}{algo:bc:upon:input:six}): }
The first rule is triggered at process $P_i$ upon \freviewdel{the reception of}\freview{receiving} a client message $m$ to be broadcast in total order. It is responsible for waiting until a batch of $B$ requests has been accumulated, \reviewdel{and then} assigning it a local sequence number, and VCBC-broadcasting it to all replicas. In more detail, process $P_i$ proceeds as follows:
\begin{itemize}[leftmargin=*,noitemsep,nolistsep]
    \item
    If the set of delivered messages $S_i$ does not contain the client message $m$, append it to the buffer $buf_i$, or ignore it otherwise (\crefrange{algo:bc:upon:input:one}{algo:bc:upon:input:two}). 

    \item
    If the size of $buf_i$ reached a threshold $B$, input $buf_i$ to a VCBC instance tagged with \textit{ID} $(i, priority_i)$, indicating that $P_i$ assigned the local priority value $priority_i$ to a proposal consisting of the current buffer contents (\crefrange{algo:bc:upon:input:three}{algo:bc:upon:input:four}).
    
    \item Increment $priority_i$, so that it can be assigned to the next proposal from $P_i$, and clear the buffer (\crefrange{algo:bc:upon:input:five}{algo:bc:upon:input:six}).
\end{itemize}

\noindent \textbf{Upon rule 2 (\crefrange{algo:bc:upon:deliver}{algo:bc:upon:deliver:four}): }
The second rule is triggered at process $P_i$ upon the delivery of a proposal $m$ for a given VCBC instance tagged with \textit{ID} $(j, priority_j)$, where $j$ corresponds to the identifier of the replica $P_j$ that proposed $m$, and $priority_j$ to the sequence number assigned to it by $P_j$. Process $P_i$ proceeds as follows:
\begin{itemize}[leftmargin=*,noitemsep,nolistsep]
    \item
    Insert the delivered proposal $m$ into the slot $priority_j$ of the priority queue $Q_j$, mapping to $P_j$\reviewdel{This corresponds to adding the $m$ to the backlog of $P_j$ in $P_i$, with the priority value $priority_j$} (\crefrange{algo:bc:upon:deliver:one}{algo:bc:upon:deliver:two}). \review{This corresponds to $P_i$ updating its view on the state of $P_j$'s pending requests.}

    \item
    If the set $S_i$ contains $m$, indicating that it had already been delivered, then process $P_i$ immediately removes it from $Q_j$ to
    \review{prevent a duplicate delivery that would violate the integrity property}\reviewdel{avoid ordering duplicate messages, effectively deleting $m$ and progressing through the queue} (\crefrange{algo:bc:upon:deliver:three}{algo:bc:upon:deliver:four}).
\end{itemize}


\begin{algorithm}[h]
\begin{algorithmic}[1]
\small
\algdef{SN}[constants]{Constants}{EndConstants}
{\textbf{constants:}}

\algdef{SN}[variables]{Variables}{EndVariables}
{\textbf{state variables:}}

\algdef{SN}[upon]{Upon}{EndUpon}
    [1][]{\textbf{upon} #1 \textbf{do}}

\Statex
\Constants\label{bc:const}
    \State $B$
\EndConstants\label{bc:const-vars}

\vspace{-7px}

\Statex
\Variables\label{bc:vars}
    \State $buf_i$\label{algo:bc:vars:buff}
    \State $priority_i$ \label{algo:bc:vars:priority}
\EndVariables\label{bc:end-vars}

\vspace{-7px}

\Statex
\Procedure{BC-START}{}
    \State $buf_i \gets \emptyset$
    \State $priority_i \gets 0$
\EndProcedure

\vspace{-7px}

\Statex
\Upon[receiving a message $m$, from a client] \label{algo:bc:upon:input}
    \If{$m \notin S_i$} \label{algo:bc:upon:input:one}
        \State $buf_i \gets buf_i \cup \{m\}$ \label{algo:bc:upon:input:two}
        \If{$|buf_i| = B$} \label{algo:bc:upon:input:three}
            \State \textbf{input} $buf_i$ to \texttt{VCBC} $(i,  priority_i)$ \label{algo:bc:upon:input:four}
            \State $buf_i \gets \emptyset$ \label{algo:bc:upon:input:five}
            \State $priority_i \gets priority_i+1$ \label{algo:bc:upon:input:six}
        \EndIf
    \EndIf
\EndUpon\label{bc:end-update}

\vspace{-7px}

\Statex
\Upon[outputting $m$ for \texttt{VCBC} $(j,  priority_j)$]\label{algo:bc:upon:deliver}
    \State $Q_j \gets queues_i[j]$ \label{algo:bc:upon:deliver:one}
    \State $Q_j.Enqueue(priority_j, m)$ \label{algo:bc:upon:deliver:two}
    \If{$m \in S_i$} \label{algo:bc:upon:deliver:three}
        \State $Q_j.Dequeue(m)$ \label{algo:bc:upon:deliver:four}
    \EndIf
\EndUpon\label{bc:end-deliver}
\end{algorithmic}
\caption{\ProtocolX\ - Broadcast Component (at $P_i$)}
\label{algo:alea:bc}
\end{algorithm}

\subsubsection{Agreement Component}
The agreement component\freviewdel{,} presented in Algorithm~\ref{algo:alea:ac}\freviewdel{,} establishes a total order among client requests. \freviewdel{This is done}\freview{Requests are ordered} through a succession of agreement rounds that iterate through the various priority queues and decide whether to insert the head of that queue in the total order or skip it.
Processes maintain a single state variable $r_i$, serving as a unique identifier for the current agreement round. The execution of the agreement component starts with a call to the \textproc{AC-Start} procedure (line~\ref{algo:ac:start}), which initializes the local variable $r_i$ to $0$ and begins executing the agreement loop.



\noindent \textbf{Agreement loop (\crefrange{algo:ac:start:loop:start}{algo:ac:start:loop:end}): }
For each iteration $r_i$ of the agreement loop, the queue of proposals pertaining to a certain replica is selected.
This replica is a designated round leader, chosen through a deterministic function of the round number $F$ (e.g., by rotating through all replicas).
Let $P_a$ denote the current round leader, \review{and $Q_a$ the corresponding priority queue at each replica $r_i$.}\reviewdel{such that all replicas operate over the priority queue $Q_a$ for $r_i$.}\abnote{é um pouco estranho introduzir esta notação ($P_a$/$Q_a$) no texto mas não no código}
Process $P_i$ proceeds as follows:
\begin{itemize}[leftmargin=*,noitemsep,nolistsep]
    
    \item
    Run an ABA instance with id $(r_i)$ to determine whether the $value$ in the head slot of $Q_a$ should be delivered in this round.
    Process $P_i$, inputs $1$ to ABA if its local $Q_a$ contained $value$ in the head slot, or $0$ otherwise (\crefrange{algo:alea:ac:start:loop:one}{algo:alea:ac:start:loop:four}).
    
    \item
    If the ABA execution decided for $0$, indicating that no proposal should be delivered for the current round $r_i$, simply proceed to the next loop iteration, otherwise:
    \begin{itemize}
        \item
        If process $P_i$ input $0$ \review{to}\reviewdel{into the} ABA, send a \texttt{FILL-GAP} message to all processes that voted for $1$. This\freview{ step} is required because, at this point in time, $P_i$ is unaware of the value to deliver for $r_i$ and, therefore, must request it from another process.
        (\crefrange{algo:ac:start:loop:fg:start}{algo:ac:start:loop:fg:end}).
        
        \item
        Block execution until the head slot of $Q_a$ contains a value to be delivered via a call to the \textproc{AC-Deliver} procedure (\cref{algo:ac:start:loop:block}). The value of the head slot can be updated by the delivery \freviewdel{by}\freview{of} a pending VCBC instance, either through ``normal'' execution or as \freview{a }result of the reception of a \texttt{FILLER} message.
    \end{itemize}
\end{itemize}

\noindent
In addition to the main agreement loop, the agreement component also defines two upon rules associated with the recovery sub-protocol to handle the reception of valid \texttt{FILL-GAP} and \texttt{FILLER} messages:

\noindent \textbf{Upon rule 1 (\crefrange{algo:ac:upon:fill}{algo:ac:upon:fill:five}):}
The first rule is triggered by any correct process $P_i$ upon the reception of a valid \fillgap{q}{s} message from $P_j$, where $q$ identifies a priority queue $Q_q$, and $s$ specifies the current head slot of $Q_q$ in $P_j$. Process $P_i$ then proceeds as follows:
\begin{itemize}[leftmargin=*,noitemsep,nolistsep]
    \item
    Check if its local queue pertaining to $P_q$ is more advanced than the one of $P_j$, by comparing the head pointer of its $Q_q$ against $s$ (\cref{algo:ac:upon:fill:three}). If it is lower, \freviewdel{this indicates that }$P_i$ cannot satisfy the \texttt{FILL-GAP} request\freviewdel{,} \freview{and }thus ignor\freviewdel{ing}\freview{es} it\freviewdel{,}\freview{.} \freviewdel{o}\freview{O}therwise:
    
    \noindent --
        Compute and store in $entries$ a verifiable message $M$\freviewdel{,} for all VCBC instances originating from $P_q$ tagged with a priority comprised between the value $s$, requested by $P_j$, and the current head slot of $Q_q$ in $P_i$ (\cref{algo:ac:upon:fill:four}).
        
        \noindent --
        Send a \texttt{FILLER} message to $P_j$ containing all the VCBC verifiable messages $M$, computed in the previous step (\cref{algo:ac:upon:fill:five}).
\end{itemize}

\noindent \textbf{Upon rule 2 (\crefrange{algo:ac:upon:filler}{algo:ac:upon:filler:two}):}
The second rule is triggered by any correct process $P_i$ upon receiving a valid \filler{entries} message. This message is received as a response to a \texttt{FILL-GAP} request. It contains the required information necessary for $P_i$ to progress in the execution of the protocol\freviewdel{,} by completing pending VCBC instances\freviewdel{,} after blocking in line~\ref{algo:ac:start:loop:block}. Process $P_i$ proceeds as follows:
\begin{itemize}[leftmargin=*,noitemsep,nolistsep]
    \item Deliver all $M$ messages in $entries$ to the corresponding VCBC instances.
    Note that the verifiability property of VCBC ensures that it immediately terminates upon the reception of $M$, therefore triggering the second upon rule of the broadcast component.
\end{itemize}

\noindent
Finally, the \textproc{AC-Deliver} procedure (line~\ref{algo:ac:deliver}), called during the execution of the agreement loop, is responsible for delivering the contents of $value$, a batch of totally ordered messages $m$, to the application layer (line~\ref{algo:ac:deliver:app}).
Additionally, this procedure also removes $value$ from all priority queues and appends its contents to the set of delivered requests $S$. 
\review{Note that if batching is naively used, this scheme would likely lead to some redundant work being done by the replicas, as large batches differing only in a few requests could not be removed from the priority queues (in line~\ref{algo:ac:deliver:remove_duplicate}), and therefore redundant operations would go through agreement and only be removed before attempting to execute them (line~\ref{algo:ac:deliver:check}). To avoid this, we steer \freviewdel{to}\freview{the} protocol towards all replicas having the same batches by having the client optimistically submit requests to a single replica. If, after a timeout, the client does not receive a response, then it resubmits to all replicas. Furthermore, a real-world implementation would place an upper bound on the number of broadcast but not delivered requests, which implies that requests are not batched as soon as they are received but instead stay in a pool until the protocol progresses. Because of this, deduplication can be made before the batch is created, avoiding the redundant work problem.} 


\begin{algorithm}[h]
\begin{algorithmic}[1]
\small

\algdef{SN}[variables]{Variables}{EndVariables}
{\textbf{state variables:}}

\algdef{SN}[wait]{Wait}{EndWait}
    [1][]{\textbf{wait until} #1 \textbf{then}}
    
\algdef{SE}[DOWHILE]{Do}{doWhile}
    {\algorithmicdo}[1]{\algorithmicwhile\ #1}%

\algdef{SN}[for each]{ForEach}{EndFor}
    [1][]{\textbf{for each} #1 \textbf{do}}

\algdef{SN}[upon]{Upon}{EndUpon}
    [1][]{\textbf{upon} #1 \textbf{do}}

\Statex
\Variables\label{ac:vars}
    \State $r_i$ \label{ac:vars:r}
\EndVariables\label{ac:end-vars}

\vspace{-7px}

\Statex
\Procedure{AC-START}{}\label{algo:ac:start}
    \State $r_i \gets 0$
    \While {true}\label{algo:ac:start:loop:start}
        \State $Q \gets queues_i[F(r_i)]$\label{algo:alea:ac:start:loop:one}
        \State $value \gets Q.Peek()$\label{algo:alea:ac:start:loop:two}
        \State $proposal \gets value \neq \bot$ ? $1 : 0$ \label{algo:alea:ac:start:loop:three}
        \State \textbf{input} $proposal$ to \texttt{ABA} $(r_i)$ \label{algo:alea:ac:start:loop:four}
        \Wait[\texttt{ABA} $(r_i)$ delivers $b$]
            \If{$b = 1$}
                \If{$Q.Peek() = \bot$}\label{algo:ac:start:loop:}\label{algo:ac:start:loop:fg:start}
                    \State broadcast $\langle \texttt{FILL-GAP}, Q.id, Q.head \rangle$\label{algo:ac:start:loop:fg:end}
                \EndIf
                \Wait [$(value \gets Q.Peek()) \neq \bot$]\label{algo:ac:start:loop:block}
                    \State \textproc{ac-deliver}$(value)$
                \EndWait
            \EndIf
        \EndWait
    \State $r_i \gets r_i+1$
    \EndWhile\label{algo:ac:start:loop:end}
\EndProcedure\label{end-procedure:start}

\vspace{-7px}

\Statex
\Upon[receiving a valid $\langle \texttt{FILL-GAP}, q, s \rangle$ message from $P_j$] \label{algo:ac:upon:fill}
     \State $Q \gets queues_i[q]$\label{algo:ac:upon:fill:one}
     \If{$Q.head \geq s$}\label{algo:ac:upon:fill:three}
        \State $entries \gets $\texttt{VCBC}$(queue, s').M$ $\forall_{s'} \in [s, Q.head]$ \label{algo:ac:upon:fill:four}
        \State \textbf{send} $\langle \texttt{FILLER}, entries \rangle$ to $P_j$\label{algo:ac:upon:fill:five}
     \EndIf
\EndUpon

\vspace{-7px}

\Statex
\Upon[delivering a valid $\langle \texttt{FILLER}, entries \rangle$ message] \label{algo:ac:upon:filler}
    \ForEach [message $M \in entries$]\label{algo:ac:upon:filler:one}
        \State \textbf{deliver} $M$ to the corresponding \texttt{VCBC}\label{algo:ac:upon:filler:two}
    \EndFor
\EndUpon

\vspace{-7px}

\Statex
\Procedure{ac-deliver}{value}\label{algo:ac:deliver}
    \ForEach [$Q \in queues_i$]
        \State $Q.Dequeue(value)$\label{algo:ac:deliver:remove_duplicate}
    \EndFor
    \ForEach [$m \in value$]
        \If{$m \notin S_i$} \label{algo:ac:deliver:check}
            \State $S_i \gets S_i \cup \{m\}$
            \State \textbf{output} $m$ \label{algo:ac:deliver:app}
        \EndIf
    \EndFor
\EndProcedure
\end{algorithmic}
\caption{\ProtocolX\ - Agreement Component (at $P_i$)}
\label{algo:alea:ac}
\end{algorithm}

\section{Optimizations}
\label{sec:opt}

As we implemented and tested \ProtocolX, we developed the following optimizations \freviewdel{for improving}\freview{to improve} its performance.

\noindent \textbf{Input unanimity.} When a replica observes all $N$ replicas providing as input the same value $v$ to an ABA instance, then it is guaranteed that the ABA will decide $v$.
To leverage this observation, we added an early termination path to the ABA protocol. This is achieved by modifying the \texttt{INIT} message (which is only sent once at the start of the protocol) to convey the input of each replica. Then, when a replica receives $N$ modified \texttt{INIT} messages with the same value $v$, it immediately delivers $v$\freviewdel{,} and broadcasts \texttt{FINISH}\freview{ (if not broadcast yet)}.
Crucially, it continues executing the ABA protocol normally until it receives $2f+1$ \texttt{FINISH} messages, as only then it is guaranteed that all correct replicas can eventually terminate.

\noindent \textbf{Pipelining prediction.} To maximize the chances of a successful outcome of the ABA stage, replicas keep statistics about the time to complete previous VCBC and ABA executions and use that information to fine-tune the pipeline and \freviewdel{also }adapt it to the network conditions.
In particular, replicas delay negative votes for an ABA when a VCBC for the slot being voted is still in progress but is expected to end soon (according to the current estimate), with the expectation that the time to complete the broadcast is smaller than the cost of a negative ABA result.
Additionally, replicas \freviewdel{also }\review{anticipate} batch formation (and consequently the start of VCBC) \review{when deemed useful} to minimize the chance of a negative ABA result.\reviewdel{while striving (but not demanding) to keep batches full.} This is achieved by attempting to time the start of the broadcast, such that it ends right before the corresponding ABA.




\reviewdel{
\textbf{Eager ABA.} When a broadcast for a slot in the head of a queue completes at a given replica, it immediately knows that its input for the corresponding ABA for that queue will be $1$.
As such, rather than waiting for the corresponding agreement round to begin, we eagerly start it with input $1$.
To avoid overloading the network, we impose two restrictions: the number of agreement rounds that can receive this early input ahead of the current round is restricted to a maximum value, and the ABA is not allowed to make progress until the agreement round begins or the unanimity condition is reached.}\reviewdel{
This, however, introduces the possibility of a future agreement round delivering before the current one. To address this, the agreement component must buffer deliveries for future agreement rounds, to guarantee that they are always delivered in order.}


\noindent \textbf{Leader prediction.} Latency can be improved if the client sends the request to a replica that is about to become a leader: if that happens, that replica will quickly include it in the next batch to be processed and delivered. In situations with low load and where a single client issues a sequence of requests, we found that using a round-robin approach is very effective because the rate of requests followed the leader rotation. \review{Alternatively, clients can receive periodic hints from the replicas about the rotation schedule or rely on the replica they contact to redirect the request to a faster replica.}


\section{Analysis}\label{sec:efficiency}

This section analyzes the asymptotic efficiency of the \ProtocolX\ protocol according to time, message, and communication complexity metrics. The results of this analysis are summarized in Table~\ref{table:alea-complexity}.

To analyze \ProtocolX\, we observe that message exchanges occur in three places for each proposal payload to be delivered.
First, during the execution of the broadcast component, a replica initiates a VCBC instance to disseminate the locally ordered proposal.
Second, all replicas participate in successive ABA executions to decide whether or not to deliver the proposal in a particular slot. Here, we denote by $\sigma$ the average number of ABA instances executed over a given slot to reach a positive decision.
Finally, a fetch request is triggered by replicas that did not VCBC-deliver the proposal before the corresponding ABA decided $1$.

\subsection{Time Complexity}
Time complexity is defined as the expected number of communication \review{steps}\reviewdel{rounds before a protocol terminates, or, in case of a continuously running protocol such as atomic broadcast,} from a client request to its output. In the case of \ProtocolX, the first and third steps terminate in constant time $\mathcal{O}(1)$. In contrast, the total number of rounds required for the agreement component to decide depends on the value of $\sigma$, therefore giving a\freview{n expected} time complexity of \ProtocolX\ of $\mathcal{O}(\sigma)$.\abnote{alguém que lê isto pode pensar que a complexidade de cada ABA é O(1) constante (em vez de O(1) expected), mas como falamos antes que é O(1) expected deve ser ok (e não temos muito espaço para expandir)}

\subsection{Message Complexity}
We measure message complexity as the expected number of messages generated by correct replicas to execute a single client request.
In \ProtocolX, the VCBC instance from the broadcast phase generates $\mathcal{O}(N)$ messages; then, every ABA instance exchanges $\mathcal{O}(N^2)$ messages\freview{ in expectation}; and finally, the third recovery phase incurs an overhead of $\mathcal{O}(N)$ messages per replica that triggers this fallback protocol.
Hence, the\freview{ expected} message complexity of \ProtocolX\ is $\mathcal{O}(\sigma N^2)$, due to the $\sigma$ ABA instances that are executed per priority queue slot \freviewdel{prior to}\freview{before} delivery, \review{which is close to the quadratic lower bound on message complexity shown by Dolev and Reischuk~\cite{dolev1985bounds}}.

\subsection{Communication Complexity}
Communication complexity consists of the expected total bit-length of messages generated by correct replicas during the protocol execution.
Let $|m|$ correspond to the average proposal size and $\lambda$ the size of a threshold signature share.
The execution of VCBC incurs a communication complexity of $\mathcal{O}(N(|m|+\lambda))$. Each ABA instance requires correct \reviewdel{node}\review{replica}s to exchange $\mathcal{O}(\lambda N^2)$ bits\freview{ in expectation}, and finally, each replica that triggers the recovery phase adds communication cost of $\mathcal{O}(N(|m|+\lambda))$ bits.
This results in an expected total communication complexity of $\mathcal{O}(N^2(|m| + \sigma\lambda))$\freviewdel{,} due to $\sigma$ ABA executions and up to $N$ recovery rounds being triggered.

\begin{table}[htb]
\centering
\normalsize
\caption{Complexity of \ProtocolX\ decomposed by stages.}
\label{table:alea-complexity}
{
    \begin{tabular}{ | c | c | c | c | } \hline
    \textbf{Stage}  & Message                   & Communication                                       & Time                  \\ \hline
    Broadcast       & $\mathcal{O}(N)$          & $\mathcal{O}(N(|m|+\lambda))$                       & $\mathcal{O}(1)$      \\ \hline
    Agreement       & $\mathcal{O}(\sigma N^2)$ & $\mathcal{O}(\sigma\lambda N^2)$                    & $\mathcal{O}(\sigma)$ \\ \hline
    Recovery        & $\mathcal{O}(N^2)$        & $\mathcal{O}(N^2(|m|+\lambda))$                     & $\mathcal{O}(1)$      \\ \hline
    \textbf{Total}  & $\mathcal{O}(\sigma N^2)$ & $\mathcal{O}(N^2(|m| + \sigma\lambda))$             & $\mathcal{O}(\sigma)$ \\ \hline
    \end{tabular}
    }
\end{table}

\subsection{Estimating $\sigma$}
As previously mentioned, \ProtocolX\ does not guarantee a constant-time execution, which could negatively affect the protocol latency. In particular, this is because multiple zero-deciding ABA instances could be executed over the same priority queue slot until its contents are considered totally ordered.
However, we argue that, despite being theoretically unbounded, the value of $\sigma$ (the number of ABA instances required for a decision) is, in practice, close to the optimal value of 1. This is justified by the observation that, in a round-robin leader assignment, each queue is revisited every $N$ rounds, meaning that $N-1$ other ABA instances were executed by the time a given queue is revisited.
Considering ABA's validity property, which states that the decided value must have been proposed by a correct process, the termination of a VCBC instance by $N-f$ correct replicas guarantees that the next ABA execution pertaining to it will decide for $1$.
Therefore, for the value of $\sigma$ to increase by a single unit, correct replicas would, on average, have to complete $N$ sequential ABA executions for every single VCBC instance.
In our experiments, we validated that the value of $\sigma$ was very close to 1 in practice.

\section{Correctness}
\label{sec:corr}

Next, we sketch the correctness of the protocol, and we provide a complete proof \reviewdel{in our arXiv report from February 2022~\cite{alea2022arxiv} and also }\review{in a separate arXiv e-print~\cite{alea2022arxiv}}. The complete proof follows the structured proof format by Lamport~\cite{lamport_proof} for increased preciseness.

\noindent \textbf{Safety.} The two safety properties that need to be proven are integrity (each
message $m$ appears at most once in the delivery sequence of correct process $i$) and total order (any two messages $m$ and $m'$ are delivered in the same order by any pair of correct processes $i$ and $j$).
Integrity \review{is derived}\reviewdel{ derives from the construction of the data structures used by \ProtocolX, namely updating the set of delivered messages after delivery and also dequeuing and never reenqueueing a message that was delivered. } \review{from the fact that once a message is delivered, it is added to the set of delivered messages, dequeued from all queues and never enqueued again.}
\reviewdel{In turn, t}\review{T}he total order property follows from \reviewdel{a contradiction proof, where}\review{the fact that} delivering two different messages at different replicas in the same slot would lead to a violation of the consistency property of VCBC.

\noindent \textbf{Liveness.} The liveness properties build mostly on the liveness guarantees of the
protocols \freviewdel{that are }used as building blocks. Given these guarantees, it suffices to follow the protocol steps \freviewdel{in order }to prove
that we eventually satisfy the preconditions for the building block protocols to
produce the necessary outputs to decide a value, namely that the messages that were broadcast reach a sufficient number of correct processes.

\noindent \textbf{Censorship resilience.} Prior asynchronous BFT protocols include mechanisms to enforce that Byzantine \reviewdel{node}\review{replica}s cannot significantly delay the delivery of any particular message (i.e., a fairness property). This is required, in particular, for protocols that use an asynchronous common subset (ACS) to agree on a subset of the various proposals from different replicas to deliver, since a Byzantine \reviewdel{node}\review{replica} can bias the choice of proposals to be included in the output of ACS. In \ProtocolX, however, censorship resilience is easily achieved by construction, given that any replica can initiate a VCBC for a client request. Thus, clients can broadcast their requests to $f+1$ or more replicas (possibly after a wait, to optimistically check if sending to a single replica suffices). This guarantees that at least one of these replicas is non-faulty and will drive the request execution.

\section{Implementations}\label{sec:impl}
We implemented \ProtocolX\ in three open-source prototypes:
an initial research implementation, and then two real-world integrations, namely with \review{the SSV} Ethereum distributed validator (where it is being considered to replace QBFT as its main protocol in the near future~\cite{ssv-blog}) and with an experimental consensus layer for \reviewdel{the subnets of a major cryptocurrency}\review{subnets in the Filecoin network~\cite{alea_mir}}.

\noindent \textbf{Research prototype.} Our first prototype implementation of \ProtocolX, which is available as open source \cite{alea_proto}, comprises $20,000$ lines of Java code.
The source code is organized in  a modular manner, with the main logic of \ProtocolX\ leveraging different subprotocols (namely broadcast and binary agreement) as building blocks. Reliable point-to-point links were implemented using TCP streams, similar to prior work~\cite{honeybadger,dumbo,dumbo-ng}. 
Additionally, we implemented HBBFT using the same codebase as a starting point to use it as one of the comparison baselines. 

\noindent \textbf{Ethereum distributed validator.}
Decentralized Validator Technology (\freviewdel{or }DVT) is a technology to improve the security, robustness, and openness of the Ethereum network~\cite{ethereum-dvt,eth-tweet}.
Using distributed validators, several non-trusted parties cooperate to logically act as a single validator, and this way each participant is able to overcome the need to commit 32 ETH (over USD $3,700$ as of this writing) to enter the network. To act as a single logical entity, once a distributed validator is called to conduct a validation task (called a duty in Ethereum), the various parties that form the validator run a BFT consensus protocol to decide on the input to the duty (which can be, for instance, a block or a pointer to the head of the chain, depending on the type of duty) and the respective outcome.

We have been collaborating with ssv.network \reviewdel{a DVT company}for over one year~\cite{ssv-blog} to implement \ProtocolX\ in the SSV codebase, with the goal of offering stronger resilience in the presence of adverse network conditions or Byzantine behavior. Their current plan is to incorporate \ProtocolX\ in their production codebase in the near future. The repository for this implementation is available as open source \cite{alea_ssv}. 


The main integration challenge came from the fact that consensus is used as a standalone instance, instead of a replicated state machine that executes a command sequence, which would be more aligned with the abstraction offered by \ProtocolX. Therefore, we adapted \ProtocolX\ with the following design features and optimizations to fit this specific context.

\noindent \textit{Adapting \ProtocolX\ to one-shot consensus.} In a distributed validator, even though the duty is \review{the same across} the processes that comprise the validator, because it is known a few epochs in advance, the input to that duty needs to be agreed upon because each process may retrieve it from a different source (called a beacon client in Ethereum, with several possible providers). To reach consensus on that input, each process will attempt to send its input using \ProtocolX's atomic broadcast protocol, and the first to be delivered by the protocol is the output of consensus. Note that only one instance of VCBC per process is needed to implement this one-shot consensus, thus simplifying the implementation. A possible concern is that the validity condition for this consensus implementation allows for a single divergent opinion to be the final output. However, this is safe because the inputs are coming from sources that are outside of the distributed validator system. Therefore their correctness is beyond the scope of the distributed validator. (In addition, some basic validation checks can also be conducted.)
Additionally, for fairness, the round-robin rotation of protocol leaders in different consensus instances is based on a pseudorandom sequence, \review{allowing the advantageous roles to even out over time}.
 
\noindent \textit{Early consensus termination.}
    As an optimization, if a \reviewdel{node}\review{replica} receives a VCBC proof for the same value for every participant, it knows in advance that the corresponding value is the only possible output of consensus and can return it immediately. However, it continues to run the consensus protocol in case other \reviewdel{node}\review{replica}s do not receive the same view. This is particularly useful in distributed validators because  \reviewdel{node}\review{replica}s have a high chance of proposing the same value. This is because, for most validations, different inputs  only occur when different \reviewdel{node}\review{replica}s have a divergent view of the current state of the blockchain, which is rare. \review{Note that this optimization differs from the first optimization described in Section~\ref{sec:opt}, which refers to ABA instead of VCBC instances.}
    



The current implementation consists of 5,000 lines of Go code, integrated as a subset of \review{the large codebase of SSV}.

\noindent \textbf{Consensus layer for \reviewdel{cryptocurrency}\review{Filecoin} subnets.}
We also implemented \ProtocolX\ as part of \review{Mir/Trantor~\cite{mir_readme,trantor}, }an experimental framework for distributed protocols, which is meant to \freviewdel{a }become \freview{a }new consensus layer for the subnets of \reviewdel{a major cryptocurrency}\review{Filecoin}~\cite{mir_readme,filecoin-hier}.
\reviewdel{This framework supports two other protocols, namely a recent proposal focused on modularity called Trantor~\cite{trantor}, and ISS~\cite{iss} (plus an upcoming implementation of a new protocol).}
\review{This framework already supports the ISS-PBFT~\cite{iss} protocol, and an upcoming implementation of a new protocol, apart from \ProtocolX.}
We implemented \ProtocolX\ in \freviewdel{4K}\freview{4,000} lines of Go code and are currently merging it into the main repository~\cite{mir_readme}. It is nonetheless already freely available as open source\cite{alea_mir}.
\reviewdel{Following the existing code organization, the subprotocols used by \ProtocolX\ -- ABA and VCBC -- were implemented as separate modules of this framework, and therefore can be reused by other protocols to be implemented in this framework.}
\review{The subprotocols used by \ProtocolX\ -- ABA and VCBC -- are implemented as independent modules, allowing their reuse by other protocols to be implemented within the same framework.}
In addition, we added support for threshold cryptography and BLS signatures within the framework, to be available to other protocols.

\review{
We improved the performance of this implementation through the parallel execution of agreement rounds.
In Alea-BFT, the system can only order one batch of requests per agreement round, capping the system throughput to $\textit{BatchSize} \times \textit{ABARate}$.
To overcome this limit, we allow multiple agreement rounds to make progress in parallel, but they buffer the delivery until it is guaranteed that the instances are delivered in order.
However, we have to be careful so that this parallelization does not overload the network, while also remaining effective.
To this end, we limit parallelization to the next $N$ agreement rounds and restrict ABA execution as follows. Before all the preceding agreement rounds deliver, ABA instances are only allowed to make progress using the unanimity optimization and otherwise need to wait for their turn.
Under this restriction, the eager execution of ABA instances only broadcasts up to two messages (INIT and FINISH), thus limiting its network impact.
}




\vspace{-0.75px}
\section{Evaluation}\label{sec:eval}
We evaluated the three implementations of \ProtocolX\ \review{under several scenarios}.
\reviewdel{, and, given our extensive set of results, we analyze the performance of the research prototype in depth, and the other two systems more succinctly.} The following questions guide our evaluation. (1) How does the performance of \ProtocolX\ compare to other asynchronous BFT protocols across different configurations? (2) How robust is \ProtocolX\ to faults? (3) How do the real-world implementations of \ProtocolX\ compare to the previous protocols that they employed?



\subsection{Experimental environment}

\noindent \textbf{Baselines.} The research prototype of \ProtocolX\ was evaluated against two baselines. First, we compared it to our own implementation of HBBFT~\cite{honeybadger}, the first protocol in the new generation of asynchronous BFT, and where, to obtain an apples-to-apples comparison, we started from the same codebase as \ProtocolX\ and tried to optimize the implementation of HBBFT to the fullest extent. Second, we compared to Dumbo-NG~\cite{dumbo-ng}, which is the state-of-the-art asynchronous BFT protocol with outstanding performance (several-fold better than the direct competitors, according to their results~\cite{dumbo-ng}), and, in this case, we deploy their unmodified codebase. For the two real-world implementations, we used the protocols that those systems originally supported as baselines. 

\noindent \textbf{Setup.} We deployed \ProtocolX\ in a cluster where the replica and client instances ran on machines equipped with AMD EPYC 7272 12-Core Processors. In addition, Docker was used to limit each replica's CPU usage to 4 cores and the Java VM was capped at 10GB of memory.
These machines were connected to the same local network with 1Gb connections. To evaluate the effects of deploying \ProtocolX\ in a wide-area network, we emulate varying additional inter-\reviewdel{node}\review{replica} latency using \texttt{netem}.
\review{Some} deployment characteristics for the real-world prototypes differed whenever noted.

Clients submit requests in an open loop, and we vary the inter-request interval and the number of clients to increase the load. The payload size of the requests and responses is 256 bytes, aligned with the size of Bitcoin transactions (as  noted and employed in previous work~\cite{honeybadger}). Each experiment runs for 2 minutes, and we repeat experiments \reviewdel{3}\review{5} times and report the average. \review{When measuring latency for Dumbo-NG, its implementation always loads the system, and therefore we would have to modify their codebase to measure the latency of an isolated request. As such, the latency measurements for Dumbo-NG represent the performance at an intermediate load generated by the implementation, where the system is not quiescent but also not as overloaded as during throughput measurements for the other protocols.}


\subsection{\review{Performance under different parameters}}
We start by \review{using our research prototype to measure} the \reviewdel{normal case}latency and throughput of \ProtocolX\ and the two baselines under different configurations and deployments, namely varying the batch size, the replication factor, and the inter-\reviewdel{node}\review{replica} latency. When the sensitivity to one of these parameters is being evaluated, the remaining ones are fixed to a batch size of 1024, $n=$4 replicas (i.e., $f=1$), \review{and LAN (minimal inter-node latency), respectively}. 

\begin{figure}[t]
    \centering
    \begin{subfigure}[b]{0.23\textwidth}
    \includegraphics[width=\columnwidth]{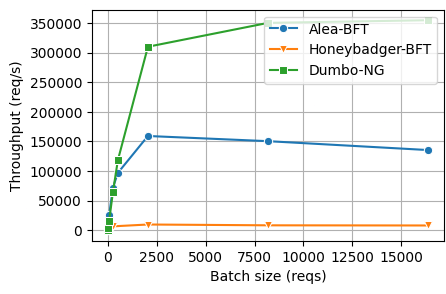}
        \vspace{-4ex}
        \caption{Peak throughput vs batch size \phantom{aaaaa}}
        \label{fig:eval:proto:varbatch:thr}
    \end{subfigure}
    \begin{subfigure}[b]{0.23\textwidth}
    \includegraphics[width=\columnwidth]{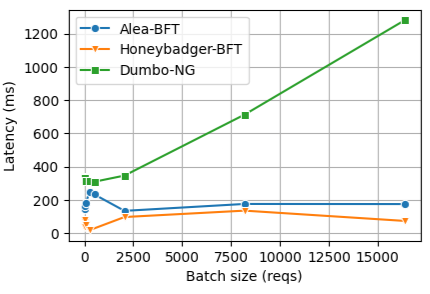}
        \vspace{-4ex}
        \caption{Latency at peak throughput vs batch size}
        \label{fig:eval:proto:varbatch:lat}
    \end{subfigure}
    
    \begin{subfigure}[b]{0.23\textwidth}
    \includegraphics[width=\columnwidth]{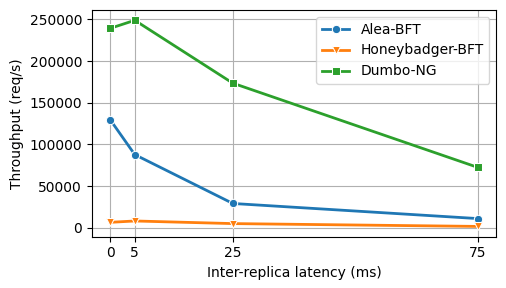}
        \vspace{-4ex}
        \caption{\reviewdel{Latency} \review{Peak Throughput vs inter-replica latency}}
        \label{fig:eval:proto:varlat:thr}
    \end{subfigure}
    \begin{subfigure}[b]{0.23\textwidth}
    \includegraphics[width=\columnwidth]{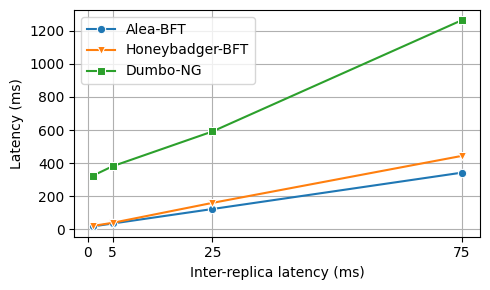}
        \vspace{-4ex}
        \caption{\reviewdel{Peak throughput} \review{Base latency vs inter-replica latency}}
        \label{fig:eval:proto:varlat:lat}
    \end{subfigure}

    \begin{subfigure}[b]{0.23\textwidth}
        \includegraphics[width=\columnwidth]{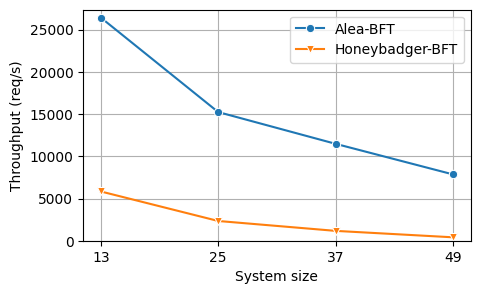}
        \vspace{-4ex}
        \caption{Peak throughput vs system size}
        \label{fig:eval:proto:scaleloaded:thr}
    \end{subfigure}
    \begin{subfigure}[b]{0.23\textwidth}
        \includegraphics[width=\columnwidth]{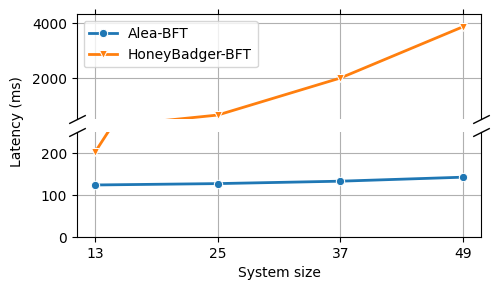}
        \vspace{-4ex}
        \caption{Base latency vs system size \phantom{aaaa}}
        \label{fig:eval:proto:scaleunloaded:lat}
    \end{subfigure}
    \begin{subfigure}[b]{0.45\textwidth}
        \includegraphics[width=\columnwidth]{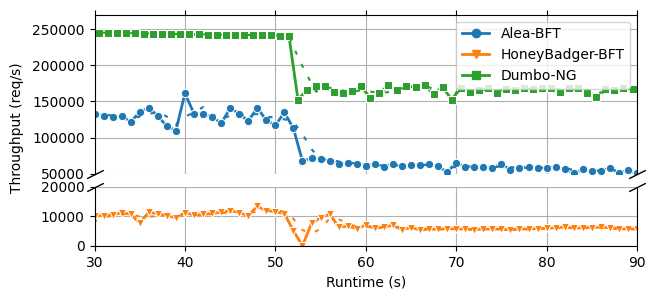}
        \vspace{-6ex}
        \caption{Throughput during crash fault\vspace{-4ex}}
        \label{fig:eval:proto:trace}
    \end{subfigure}
    
    \caption{Prototype implementation evaluation\vspace{-5ex}}
    \label{fig:eval:proto:overall}
    \centering
    \vspace{-2ex}
\end{figure}

Figures~\ref{fig:eval:proto:varbatch:thr} and~\ref{fig:eval:proto:varbatch:lat} show the peak throughput and latency while varying the batch size. \review{In this particular setting, the latency at the peak throughput is reported instead of the base latency (measuring an isolated request without any system load). This is because the latter would require issuing only one or a few requests, much less than the batch size. Thus, the measured latency would correspond to either slowly filling up the batch or triggering a batch timeout.}  The results show that \ProtocolX\ is competitive with the state of the art (Dumbo-NG) and that both are a significant improvement over their predecessor (HBBFT).
While \ProtocolX\ cannot match the peak throughput of Dumbo-NG, they are \review{both} in the same order of magnitude (hundreds of thousands of txs/s), versus $\approx$15k txs/s for HBBFT.
When considering latency, \ProtocolX\ outperforms Dumbo-NG \review{across} \reviewdel{, and can maintain a comparatively lower latency in}all tested batch sizes.
We attribute these differences in throughput and latency to the choice of agreement primitive.
\reviewdel{On the one hand, Dumbo-NG's MVBA can deliver many batches at once, but it is more expensive, making large batch sizes unnecessary and detrimental to performance. On the other hand, the simplicity of \ProtocolX's ABA  grants it a latency advantage. At the same time, the efficient VCBC contributes to pushing throughput to be on the same order of magnitude as Dumbo-NG.}
\review {In particular, Dumbo-NG uses MVBA, which allows for better throughput by accepting several batches at once, but the simpler ABA primitive of \ProtocolX\ enables a better latency under a comparable load.}
\review{Note that the fact that \ProtocolX's throughput peaks earlier than Dumbo-NG's in this setting is mainly an implementation artifact -- our codebase has an external open loop client that further saturates the network, unlike Dumbo-NG's.}


Next, we use \texttt{netem} to evaluate the performance under different network conditions (LAN vs.\ WAN).
Figures~\ref{fig:eval:proto:varlat:thr} and~\ref{fig:eval:proto:varlat:lat} show the peak throughput and latency \reviewdel{measured}when varying inter-\reviewdel{node}\review{replica} latency. \review{The results show that \ProtocolX\ has the lowest latency of all protocols while achieving a peak throughput in the tens of thousands of requests per second when the inter-node latency is under 25ms.} 
HBBFT witnesses a similar degradation in latency to Alea-BFT because the critical path for a normal-case request execution is the same for both protocols, except for a single protocol step, which explains the slightly higher latency of HBBFT. 

\begin{figure}[t]
    
\end{figure}

Finally, we scale out the experiments by increasing the number of \reviewdel{node}\review{replica}s participating in the consensus.
For these experiments, we use 13, 25, 37 and 49 replicas, and, in this case, we use \texttt{netem} to simulate a WAN environment, with a 75ms inter-\reviewdel{node}\review{replica} latency, corresponding to an RTT of 150ms (approximating a cloud deployment). \review{Furthermore,
since the available setup forced some replicas to be co-located on the same machine, to ensure a realistic and uniform bandwidth availability, each instance's bandwidth was capped at 50Mb/s using a token bucket filter.}

In this case,  we were not able to configure the Dumbo-NG code to use the same replica group sizes as the ones we employed for the other two systems, which explains why there are only two curves. 
Our results show that \ProtocolX\ not only has superior throughput but also achieves very good latency in unloaded scenarios, \review{due to the clients being able} to predict the current leader and send requests to the replica that will drive the decision the fastest. On the other hand, in unloaded scenarios, HoneyBadger's clients need to contact $2f+1$ replicas to ensure progress, meaning that, for a single \reviewdel{transaction}\review{request} to go through, $2f+1$ ABAs need to be executed.
\reviewdel{Our results show that \ProtocolX\ not only has superior throughput, but it manages to hold on to its low latency as the replica group size grows, with an important contribution from clients being able to predict who is the current leader and sending requests to that replica for faster execution.}


\subsection{Performance under faults}

We compared the performance of \ProtocolX\ and the baselines in a scenario \review{where one of the replicas crashes 50 seconds into the trace.}\reviewdel{where $f$ replicas are faulty.} We inject a crash fault instead of a protocol-specific Byzantine fault, for a direct comparison between protocols. 
The results in Figure~\ref{fig:eval:proto:trace} show that \ProtocolX\ and HBBFT suffer more with the crash of  $f$ replicas because they share the unanimity optimization described in Section~\ref{sec:opt} (which cannot be used when a replica is unresponsive), but Dumbo-NG also takes a significant hit (around 30\% of throughput) due to bandwidth wasted on the faulty replica.  
\reviewdel{Despite this}


\review{We also evaluate the performance under faults for the other implementations and present the results in the next section.}


\subsection{Real-world implementations}

\noindent \textbf{Ethereum distributed validator.}
We start this part of the evaluation with the implementation \review{of \ProtocolX\ in the SSV} distributed validator of Ethereum. In this case, the experimental methodology is constrained by the way that SSV operates, namely that the system progresses in a sequence of slots of fixed duration (12 seconds in Ethereum), and during each slot, each validator is assigned a set of duties (tasks such as block proposal and attestation). Thus, to measure the base latency, we set the number of duties per slot to 1 and measure the time to complete it, whereas throughput is measured by increasing the number of duties per slot until this metric peaks. \review{Since the performance is not network-bound and the number of nodes is low, we did not use the bandwidth cap.}
We used a group of 4 replicas, with no added inter-\reviewdel{node}\review{replica} latency, and batching is not applicable in this setting. We tested variants of \ProtocolX\ that use different message authentication methods and compared these to the existing QBFT-based codebase. In particular, we start with a direct comparison to QBFT (which, \review{in the SSV codebase}, uses BLS digital signatures without aggregation), and then add BLS  aggregation to \ProtocolX. \reviewdel{which could also be applied to QBFT}\review{(This change could also be applied to QBFT, but this was avoided to keep the baseline as the existing codebase.) Then, we replace} signatures with HMACs, which is possible in \ProtocolX\ but would not be directly applicable to QBFT, because of messages conveyed to all processes during round changes. In the case of HMACs, BLS is only used to verify the final VCBC signature and compute ABA's shared coin.



\begin{figure}[t]
    \centering
    \vspace{-1ex}
    \begin{subfigure}[b]{0.22\textwidth}
        \includegraphics[width=\columnwidth]{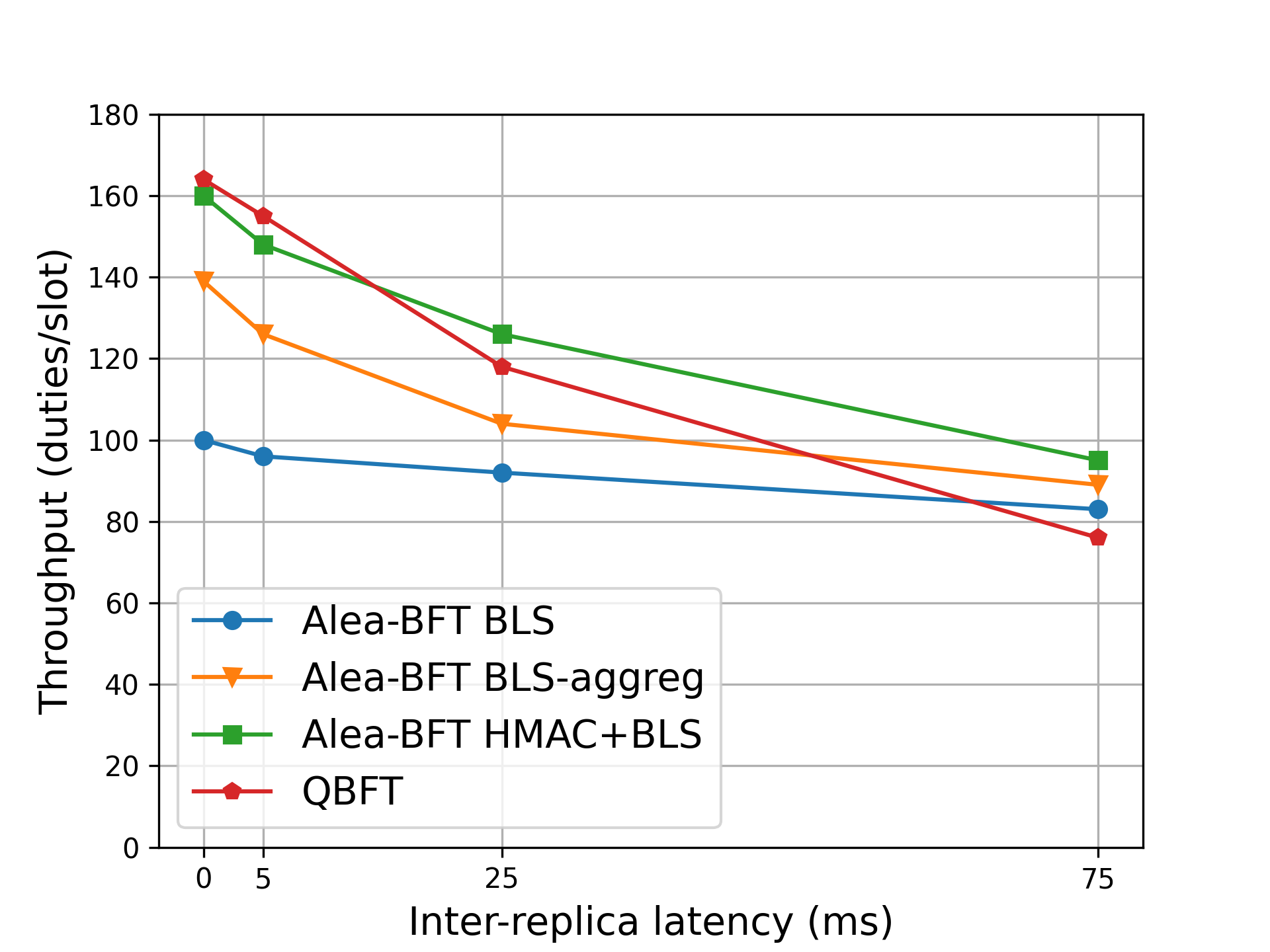}
        \caption{Peak throughput vs inter-replica latency}
        \label{fig:eval:ssv:internode_latency_throughput}
    \end{subfigure}
    \begin{subfigure}[b]{0.22\textwidth}
        \includegraphics[width=\columnwidth]{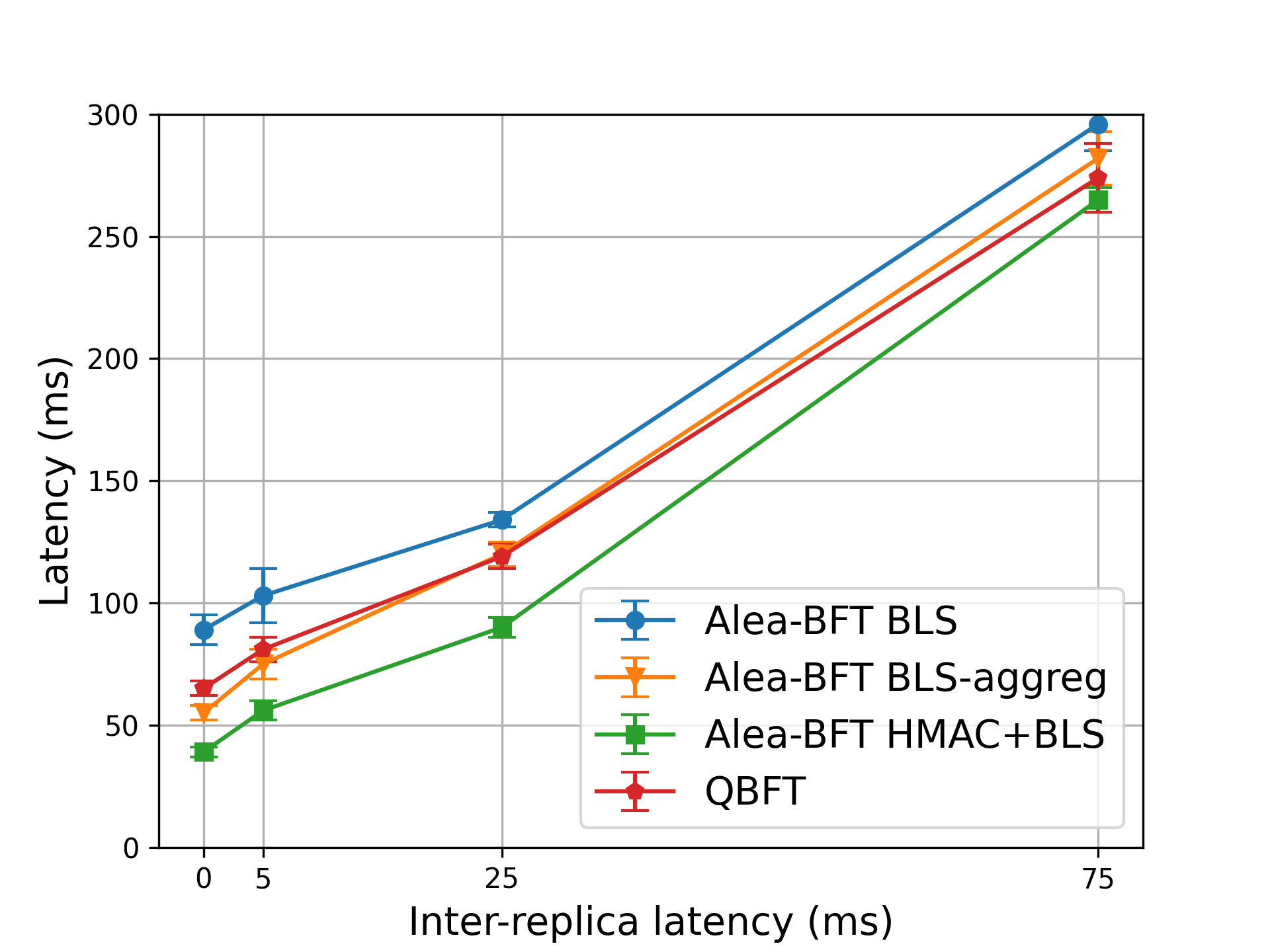}
        \caption{Base latency vs inter-replica latency}
        \label{fig:eval:ssv:internode_latency}
    \end{subfigure}
    \\
    \begin{subfigure}[b]{0.22\textwidth}
        \includegraphics[width=\columnwidth]{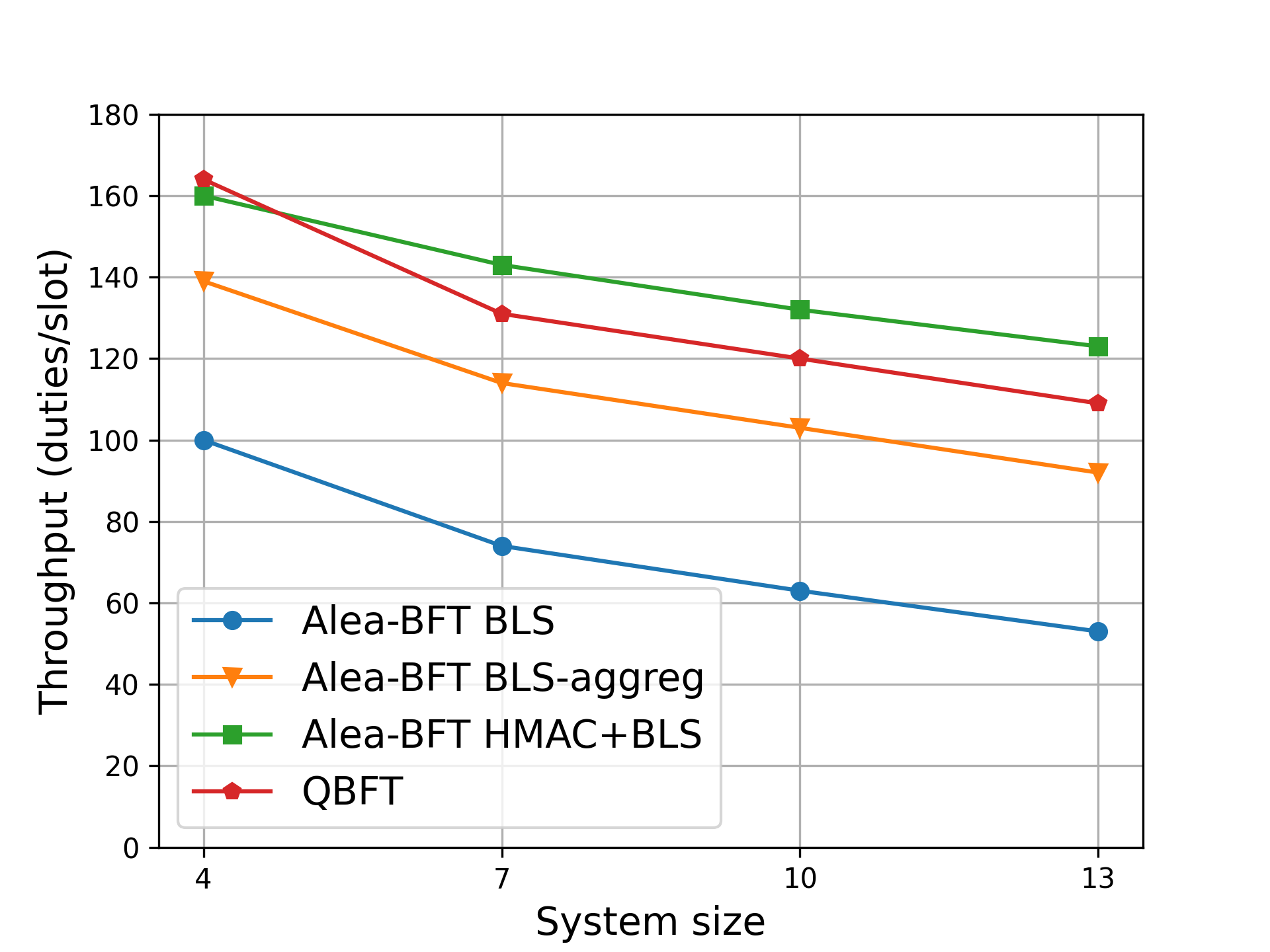}
        \caption{Peak throughput vs system size}
        \label{fig:eval:ssv:scalability_throughput}
    \end{subfigure} 
    \begin{subfigure}[b]{0.22\textwidth}
        \includegraphics[width=\columnwidth]{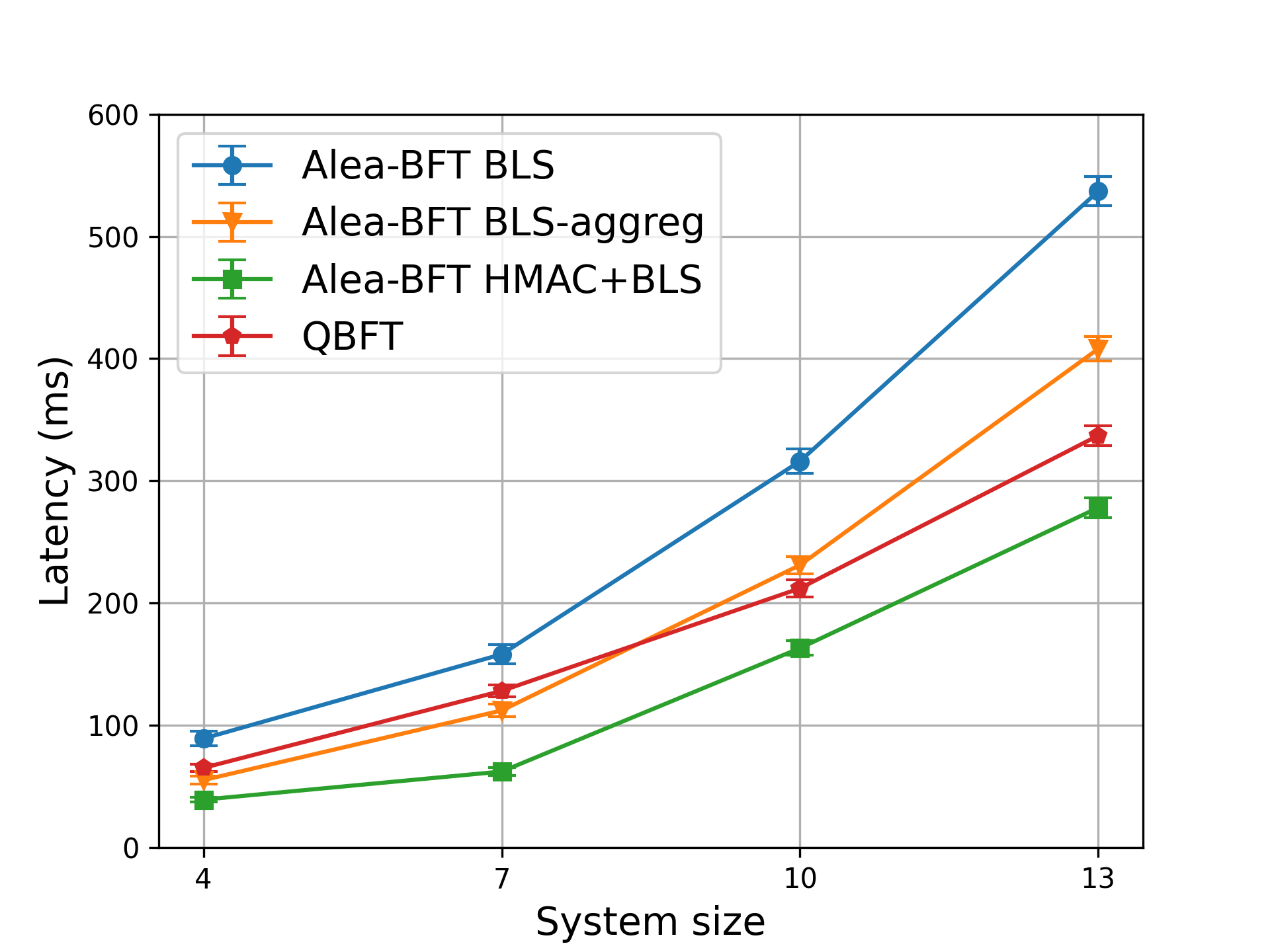}
        \caption{Base latency vs system size\phantom{asdfa}}
        \label{fig:eval:ssv:scalability}
    \end{subfigure} 
    \\
    \begin{subfigure}[b]{0.45\textwidth}
        \includegraphics[width=\columnwidth]{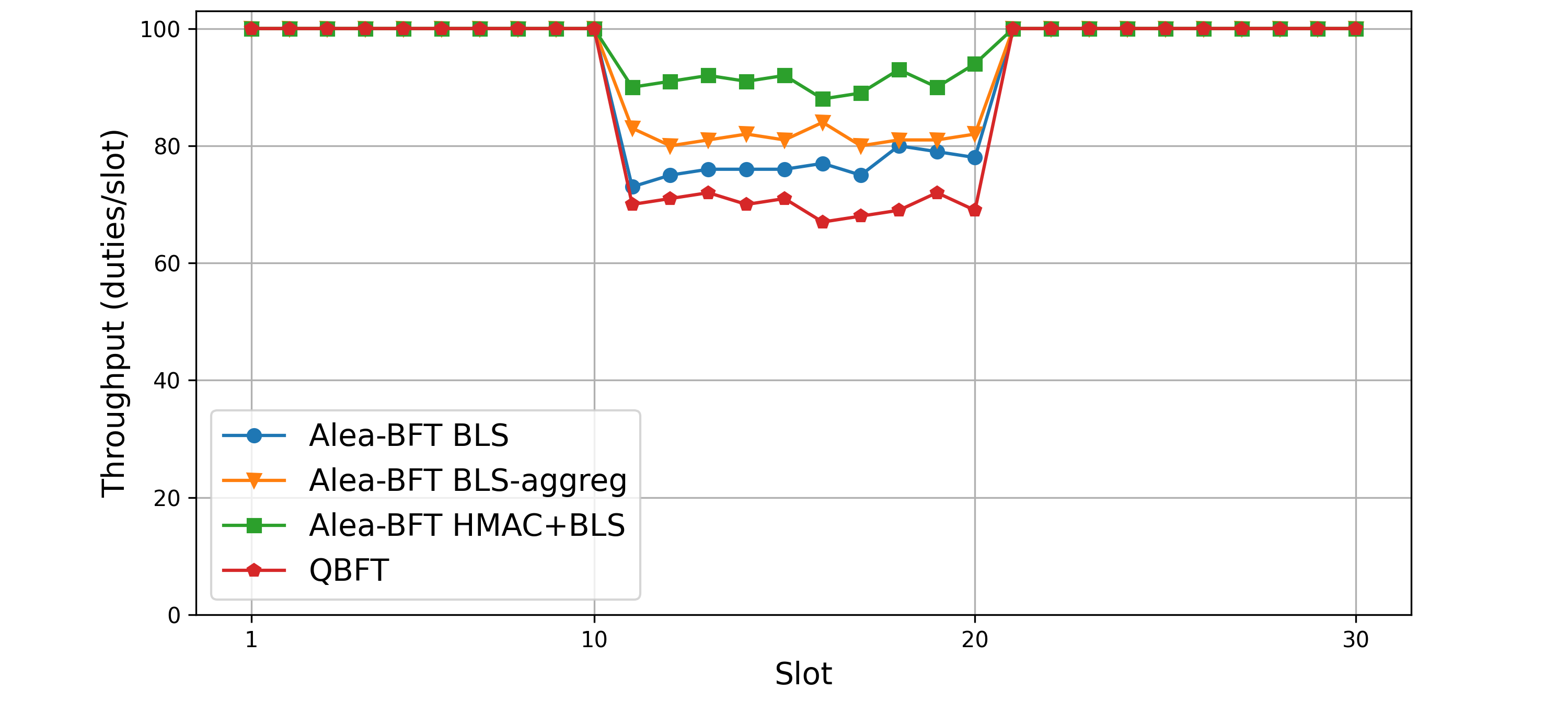}
        \vspace{-4ex}
        \caption{Throughput during crash fault}\vspace{-0.8em}
        \label{fig:eval:ssv:trace}
    \end{subfigure}
    
    \caption{Distributed validator deployment \review{evaluation}\reviewdel{with 4 replicas}\vspace{-1.5em}}
    \label{fig:eval:ssv:latency_and_throughput}
    \centering
    \vspace{-4px}
\end{figure}


Figure~\ref{fig:eval:ssv:latency_and_throughput} shows the \reviewdel{latency and  throughput of}\review{performance of} the SSV validator using different protocols. 
In these plots, the latency and throughput in the most basic setting can be determined by the leftmost point of Figures~\ref{fig:eval:ssv:internode_latency_throughput} and~\ref{fig:eval:ssv:internode_latency}. These results show that \ProtocolX\ with BLS aggregation and with HMACs has \review{similar peak throughput and} better latency than the previous codebase that uses QBFT. This highlights how designing \ProtocolX\ to have a small number of protocol steps, combined with the possibility of using HMACs that comes from not having the view change mechanism from partially synchronous protocols can lead to competitive performance in this setting. \review{We attribute the slightly better throughput of QBFT}\reviewdel{In terms of throughput, QBFT achieves a better result, which we attribute} to the leader-driven protocol allowing for exchanging a smaller number of messages. However, we see the overhead of \ProtocolX\ as a modest price to pay for not making partial synchrony assumptions.

\review{The effects of varying the inter-replica latency is shown in the remainder of Figures~\ref{fig:eval:ssv:internode_latency_throughput} and~\ref{fig:eval:ssv:internode_latency}. The key takeaway is that, for all tested conditions, \ProtocolX\  closely follows QBFT in terms of base latency and peak throughput and, with the best choice of cryptographic primitives in place, \ProtocolX\ can even achieve lower latency values. Figure~\ref{fig:eval:ssv:internode_latency} also depicts the change of relative importance of cryptographic primitives as inter-replica latency varies -- in a LAN environment, as the network delay is small relative to the cost of cryptography, the several variants have a very noticeable relative difference among them. However, as the inter-replica delay increases, this difference decreases in proportion.

Next, we measured performance as the group size increases (Figures \ref{fig:eval:ssv:scalability_throughput} and \ref{fig:eval:ssv:scalability}). Currently, in SSV, a validator can only employ 4, 7, 10, or 13 operators, as defined in its smart contract. In this experiment, as in the previous one, \ProtocolX's latency and throughput follow QBFT's, achieving lower latency  and higher throughput values when using HMAC for point-to-point authentication and BLS digital signatures.
}

Finally, Figure~\ref{fig:eval:ssv:trace} shows the results of an experiment where we crash one of the processes, chosen at random, at the beginning of the 11th slot in the run, then restart it in the 21st slot, and we plot the number of duties that are executed per slot throughout the trace. The results show that \ProtocolX\ is more resilient to this fault because of the principles behind its design: the fault will affect $1/4$ of the VCBC instances, but these rounds will quickly be skipped and replaced with productive work led by the other replicas.   
In contrast, QBFT waits for a timeout and a leader change protocol to complete, which slows down the entire system for that duration.

\noindent \textbf{Consensus layer for Filecoin subnets.}
\review{In the last part of this section}, we \reviewdel{present the preliminary evaluation of}\review{evaluate} \ProtocolX's integration into \reviewdel{a future}\review{an experimental} consensus layer~\cite{mir_readme,trantor} for \reviewdel{cryptocurrency}\review{Filecoin} subnets~\cite{filecoin-hier} \review{against the existing implementation of ISS-PBFT~\cite{iss} in the same codebase}.
\reviewdel{These experiments ran on Intel Core i5 7500 CPU machines with 16GB of RAM, connected to the same local network.
We used $f=1$}ISS-PBFT uses the same parameters as in its original evaluation~\cite{iss}.

\review{Base latency experiments use two co-located closed-loop clients per replica, ensuring incoming request queues are never empty. This was required because the implementation of ISS-PBFT stalls when there are no requests to propose. However, it has the downside of generating some load that may negatively affect latency.
We only present results for replica 0 in the ISS-PBFT latency measurements due to an implementation issue that inflates latencies in other replicas, which would unfairly harm the baseline's performance.
Peak throughput is measured in a configuration of $8*B$ closed-loop clients co-located with each replica, where $B$ is the batch size, which we empirically determined to maximize the throughput.}

\reviewdel{
We began by evaluating the protocol performance when working with a batch size of 1, to gain a general understanding of their differences.
In Figure
, we present throughput as a function of induced load. Similarly to Dumbo-NG, \ProtocolX\ underperforms the baseline in throughput but outperforms it in latency ($6.7$ms vs $12.2$ms).
}
\reviewdel{
Note that when measuring base latency, it was not enough to consider the latency of an unloaded system, as ISS does not achieve its best possible latency without a high enough load.
In each ISS epoch, every leader either delivers a fixed number of batches or delivers a smaller amount (possibly 0) and times out.
Therefore, when a leader does not have anything to propose and deliver, it will stall until it times out, and stall the rest of the system with it.
This limitation is known by the authors of ISS and the stall time is bounded by the propose timeout.
In contrast, \ProtocolX\ performs the same regardless of the base workload of any given process.
}

\begin{figure}[t]
    \centering
    \begin{subfigure}[b]{0.23\textwidth}
        \includegraphics[width=\columnwidth]{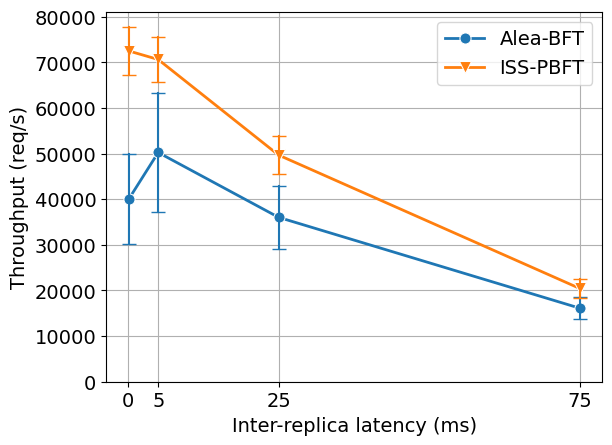}
        \vspace{-4ex}
        \caption{\reviewdel{Throughput v load ($B = 1$)}\review{Peak throughput vs inter-replica latency}}
        \label{fig:eval:mir:inl:thr}
    \end{subfigure}
    \begin{subfigure}[b]{0.23\textwidth}
        \includegraphics[width=\linewidth]{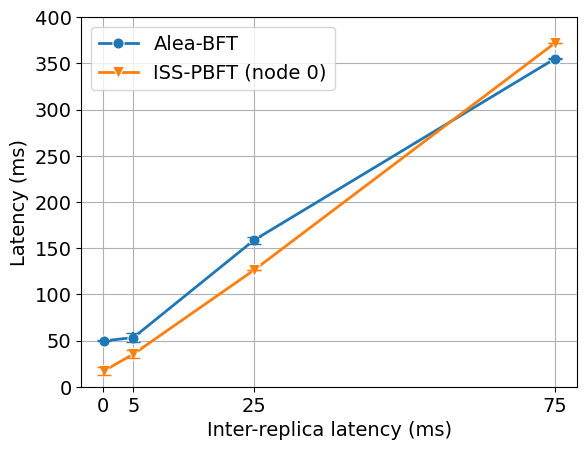}
        \vspace{-4ex}
        \caption{\reviewdel{Throughput v load ($B=2048$)}\review{Base latency vs inter-replica latency}}
        \label{fig:eval:mir:inl:lat}
    \end{subfigure}
    
    
    \begin{subfigure}[b]{0.23\textwidth}
        \includegraphics[width=\columnwidth]{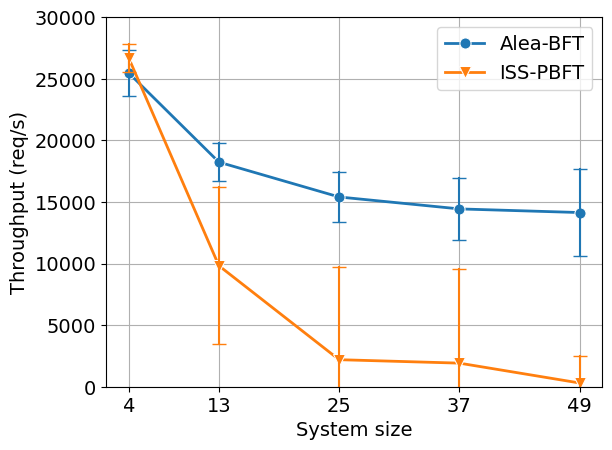}
        \vspace{-4ex}
        \caption{\review{Peak throughput vs system size}}
        \label{fig:eval:mir:scale:thr}
    \end{subfigure}
    \begin{subfigure}[b]{0.23\textwidth}
        \includegraphics[width=\linewidth]{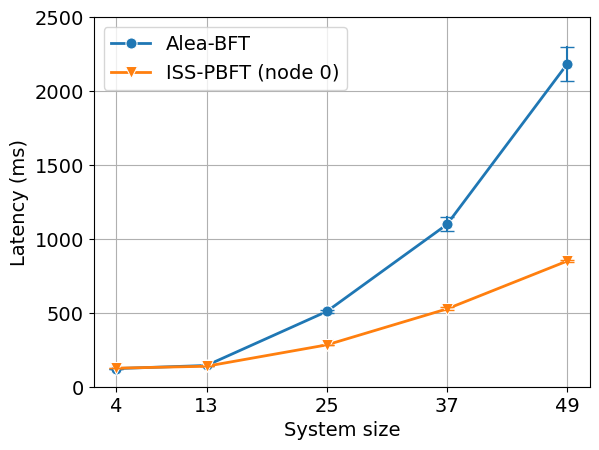}
        \vspace{-4ex}
        \caption{\review{Base latency vs system size \phantom{aaaaa}}}
        \label{fig:eval:mir:scale:lat}
    \end{subfigure}
    
     \begin{subfigure}[b]{0.45\textwidth}
        \includegraphics[width=\linewidth]{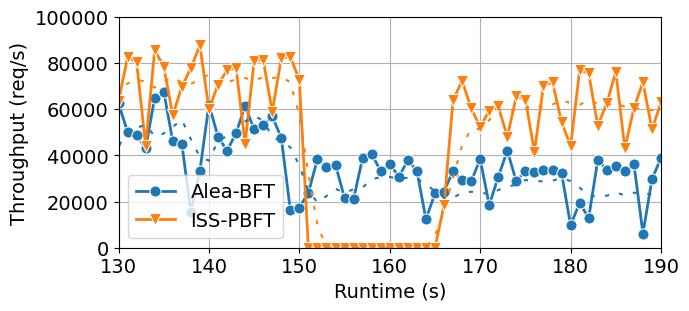}
        \vspace{-4ex}
        \caption{\review{Throughput during crash fault}\vspace{-1.6em}}
        \label{fig:eval:mir:crash}
    \end{subfigure}
    
    \vspace{1em}
    \caption{\review{Mir/Trantor deployment evaluation}}
    \label{fig:eval:mir:scale}
    \centering
    \vspace{-17px}
\end{figure}

\review{
We begin by evaluating \ProtocolX's performance against ISS-PBFT when varying the inter-node latency.
Figures \ref{fig:eval:mir:inl:thr} and \ref{fig:eval:mir:inl:lat} show that \ProtocolX\ closely follows the performance ISS-PBFT in wide-area settings, in terms of peak throughput and base latency.
Furthermore, while \ProtocolX\ is initially limited to $\approx$40k requests/s and $\approx$50ms base latency, it becomes on par with ISS-PBFT when the limiting factor shifts from threshold cryptography to  network latency.
}

\reviewdel{
Finally, we studied the impact of increasing the batch size (and induced load) to a more realistic value, namely the one used to evaluate ISS originally.
Figure
plots throughput against induced load once again, but with a batch size of 2048. Despite this \ProtocolX\ implementation not being as optimized as the more mature ISS code, our results show promise as the performance of \ProtocolX\ matches or closely follows ISS.
}

\review{
Additionally, we evaluated \ProtocolX's ability to scale against ISS-PBFT, which is relevant given that scalability is a key design goal in ISS.
Figures \ref{fig:eval:mir:scale:thr} and \ref{fig:eval:mir:scale:lat} show peak throughput and base latency measurements for \ProtocolX\ and ISS-PBFT for a variety of system sizes.
Regarding peak throughput (Figure~\ref{fig:eval:mir:scale:thr}), \ProtocolX's throughput degrades gracefully as the system size increases, fully saturating the (bandwidth-capped) network.
In contrast, ISS-PBFT degrades abruptly and stops processing requests altogether after a few seconds for $N=49$.
However, we believe this is an artifact of this implementation of ISS-PBFT, which reacts poorly under strained network conditions and is not intrinsic to the ISS-PBFT protocol.
Regarding latency (Figure~\ref{fig:eval:mir:scale:lat}), both protocols maintain near-constant base latency under system sizes up to $N=13$, after which it begins to increase.
In this case, ISS-PBFT has a lower latency than \ProtocolX\ because its multi-leader design allows requests to be processed as soon as they reach the PBFT primary replica, whereas in \ProtocolX\ we have to wait for the designated replica's turn to run its agreement round.
}


\review{
Finally, we studied the impact of crash faults on both \ProtocolX\ and ISS-PBFT.
\Cref{fig:eval:mir:crash} shows an execution trace of one \ProtocolX\ and one ISS-PBFT execution with the default settings, where a single replica crashes after 150s (and stays crashed).
To aid evaluation, a dotted line was added to both curves, showing a moving average of the system's throughput across all repetitions.
In this trace, we first observe a 15-second stall of ISS-PBFT after the crash, waiting for a timeout for the detection of  the crashed replica, whereas \ProtocolX\ can continue uninterrupted (albeit at reduced throughput) thanks to its leaderless design.
After this timeout expires, ISS excludes the crashed replica from the set of leaders and continues with a relatively small ($\approx$20\%) performance hit.
However, \ProtocolX\ is penalized on two fronts -- it both loses a replica proposing requests (like ISS) and the ABA unanimity optimization -- leading to a reduction in throughput when compared to the system with all replicas functional.
}

\section{Conclusion}\label{sec:conclusion}

In this paper, we presented \ProtocolX, a practical asynchronous BFT protocol with a design that combines simplicity with performance. 
Our experimental evaluation shows that \ProtocolX\ performs better than the top-performing Dumbo-NG in latency, offers comparable throughput, and is resilient to faults. Importantly, \ProtocolX\ is being adopted in the real\freviewdel{-}\freview{\ }world, namely by Ethereum distributed validators.

\section*{Acknowledgments}
\review{We thank the anonymous reviewers and our shepherd, Zhaoguo Wang, for their helpful feedback.
This work was supported by Fundação para a Ciência e a Tecnologia,  projects UIDB/50021/2020 and PTDC/CCI-INF/6762/2020, and by the European Union’s Horizon 2020 research and innovation programme, under grant agreement No 952226, project BIG.
}

%



\expandafter\def\expandafter\UrlBreaks\expandafter{\UrlBreaks
  \do\a\do\b\do\c\do\d\do\e\do\f\do\g\do\h\do\i\do\j%
  \do\k\do\l\do\m\do\n\do\o\do\p\do\q\do\r\do\s\do\t%
  \do\u\do\v\do\w\do\x\do\y\do\z\do\A\do\B\do\C\do\D%
  \do\E\do\F\do\G\do\H\do\I\do\J\do\K\do\L\do\M\do\N%
  \do\O\do\P\do\Q\do\R\do\S\do\T\do\U\do\V\do\W\do\X%
  \do\Y\do\Z\do\*\do\-\do\~\do\'\do\"\do\-}%
\mathchardef\UrlBreakPenalty=0
\mathchardef\UrlBigBreakPenalty=0
\bibliographystyle{plain}
\bibliography{main}

\clearpage

\appendix
\section{Correctness Proof}

\par Throughout the proofs, we assume that the instructions in the body of the upon rules described in the broadcast component are executed atomically. Atomicity can easily be achieved by using adequate synchronization primitives (such as mutexes), which were omitted to improve the readability of the protocol.

For brevity, when an assertion is made about a correct process $P_j$, its proof mostly omits references to this assumption despite requiring that $P_j$ follows the protocol.
Furthermore, any assertion \say{at the start of round $r$} or \say{at the beginning of round r} will implicitly refer  the instant before the first line of the round $r$ of Alea's agreement component for the replica that the assertion is made. \\

\textbf{Definition 1}: Let $S_i^r$  be the set of messages delivered by $P_i$ at rounds $r' < r$ when $P_i$ enters round $r$. \rrnote{, at a given point in the execution where $P_i$ has already entered round $r$ or a  higher round.} We say that consensus holds on entry at $r$ if, for any two correct processes $P_i$ and $P_j$, $S_i^r = S_j^r$. If consensus holds on entry at $r$, it's denoted $holds(r) = True$, otherwise
$holds(r) = False$. \\

Note that rounds are executed sequentially, one at a time and thus any messages delivered at rounds $r' < r$ by some process $P_j$ are delivered before round $r$. For this reason, $S_i^r$ is well-defined. \rrnote{Estou na dúvida entre dizer que é "well defined" ou que é uma "stable property" -- que achas?} \\

Let $output(VCBC_i^s, j)$ be the value delivered to $P_j$ by $VCBC_i^s$. If nothing was delivered yet, then we define $output(VCBC_i^s, j) = \bot$. Let $output(ABA_r, j)$ be the value delivered to $P_j$ by $ABA_r$. If nothing was delivered yet, we define $output(ABA_r, j) = \bot$. \\

\textbf{Definition 2 (choice function)}: Let $P_j$ be a correct process. $P_j$'s choice for $i$ at $r$ is a slot $s$ of $Q_i$ that satisfies the following conditions:

\begin{enumerate}
    \itemsep0em
    \item $\forall s' < s, output(VCBC_i^{s'}, j) \ne \bot$
    \item $\forall s' < s, output(VCBC_i^{s'}, j) \in S_j^r$
    \item $output(VCBC_i^s, j) \ne \bot$
    \item $output(VCBC_i^s, j) \notin S_j^r$
\end{enumerate}

If there's a slot $s$ that satisfies all conditions, it's denoted $s = choice_j(i, r)$.
If there is no slot that satisfies all conditions, $P_j$ is said to have no choice for $i$ at $r$ and $choice_j(i, r) = \bot$. \\

\textbf{Lemma 1}: For any $P_j$ correct, round $r$ and queue $Q_i$, $choice_j(i, r)$ is unique.

\rrnote{O que é o "suffices assume" no início da prova?}
\dsanote{acho que é engano}

\begin{proof}
\step{}{
    \assume{
        \begin{pfenum}
            \item \review{$s = choice_j(i, r)$} \rrnote{s, s' estão unbound, falta dizer que supões que existem $s=choice...$ e $s'=choice...$} \dsanote{solve I think}
            \item \review{$s' = choice_j(i,r)$}
            \item $s < s'$
        \end{pfenum}
    }
    \prove{False}
}
\step{}{$output(VCBC_i^s, j) \notin S_j^r$}
\begin{proof}
    \pf By assumption 1.1 and Definition 2 (property 4) 
\end{proof}

\step{}{$output(VCBC_i^s, j) \in S_j^r$}
\begin{proof}
    \pf By assumption 1.2 and 1.3 and Definition 2 (property 2)
\end{proof}

\step{}{$output(VCBC_i^s, j) \notin S_j^r \wedge output(VCBC_i^s, j) \in S_j^r$}
\begin{proof}
    \pf By 2 and 3
\end{proof}

\qedstep{}{
    \begin{proof}
        \pf Contradiction is reached in 4, so by 1 it's proved that the choice is unique
    \end{proof}
}
\end{proof}

Lemma 1 proves that $choice_j$ is a well-defined function. \\

\textbf{Definition 3 (value function)}: Let $P_j$ be a correct process. $value_j$ is the function that is defined as follows:

\[
value_j(i, r) = \left\{
\begin{array}{ll}
      \bot & , choice_j(i, r) = \bot \\
      output(VCBC_i^{choice_j(i, r)}, j) & , choice_j(i, r) \ne \bot\\
\end{array}
\right.
\] \\

\textbf{Lemma 2}: Let $P_j$ and $P_k$ be two correct processes and $holds(r) = True$. Let $s$ = $choice_j(i, r)$, $s'$ = $choice_k(i, r)$, $m$ = $value_j(i, r)$ and $m'$ = $value_k(i, r)$.
Then $(s \ne \bot \wedge s' \ne \bot) \implies (s = s' \wedge m = m')$.

\begin{proof}
    \step{}{
        \assume{
            \begin{pfenum}
                \item $P_j$ is correct
                \item $P_k$ is correct
                \item $holds(r) = True$
                \item $s = choice_j(i, r) \wedge m = value_j(i, r)$
                \item $s' = choice_k(i, r) \wedge m' = value_k(i, r)$
                \item $s \ne \bot$
                \item $s' \ne \bot$
            \end{pfenum}
        }
        \prove{$s = s' \wedge m = m'$}
    }
\step{}{$S_j^r = S_k^r$}
\begin{proof}
    \pf By assumptions 1.1, 1.2, 1.3 and Definition 1 
\end{proof}

\step{}{$s$ = $s'$}
\begin{proof}
    \step{}{
        \sassume{\begin{pfenum}
            \item $s < s'$
        \end{pfenum}}
        \prove{False} 
    }

    \step{}{Let $m''$ = $output(VCBC_i^s, k)$. Then $m'' \ne \bot$}
    \begin{proof}
        \pf By Definition 2 (property 1), and assumptions 1.2, 1.5, 1.7 and 3.1.1
    \end{proof}

    \step{}{$m \ne \bot$}
    \begin{proof}
        \pf By Definitions 2 (property 3) and 3, and assumptions 1.1, 1.4 and 1.6
    \end{proof}
    
    \step{}{$m$ = $m''$}
    \begin{proof}
        \pf By VCBC's consistency property, steps 3.2 and 3.3, assumptions 1.1, 1.2 and 1.4, and Definition 3
    \end{proof}

    \step{}{$m'' \in S_k^r$}
    \begin{proof}
        \pf By 3.2, Definition 2 (property 2), and assumptions 3.1.1, 1.2, 1.5 and 1.7
    \end{proof}
    
    \step{}{$m \in S_k^r$}
    \begin{proof}
        \pf By 3.4 and 3.5  
    \end{proof}
    
    \step{}{$m \in S_j^r$}
    \begin{proof}
        \pf By 3.6 and 2
    \end{proof}
    
    \step{}{$m \notin S_j^r$}
    \begin{proof}
        \pf By Definition 2 (property 4), and assumptions 1.1, 1.4 and 1.6
    \end{proof}
    
    \step{}{$m \notin S_j^r \wedge m \in S_j^r$}
    \begin{proof}
        \pf By 3.7 and 3.8
    \end{proof}

    \qedstep{}{
        \begin{proof}
            \pf By contradiction reached in 3.9 and step 3.1.
        \end{proof}
    }
\end{proof}
    
\step{}{$m$ = $m'$}
\begin{proof}
    \step{}{$m = output(VCBC_i^s, j)$}
    \begin{proof}
        \pf By Definition 3 and assumptions 1.1, 1.4 and 1.6
    \end{proof}
    
    \step{}{$m' = output(VCBC_i^s, k)$}
    \begin{proof}
        \pf By Definition 3, assumptions 1.2, 1.5 and 1.7, and step 3
    \end{proof}

    \qedstep{}{
        \begin{proof}
            \pf By VCBC's consistency property, assumptions 1.1 and 1.2, and steps 4.1 and 4.2 
        \end{proof}
    }
\end{proof}

\qedstep{}{
    \begin{proof}
        \pf By 1, 3, and 4        
    \end{proof}
}
\end{proof}

$F$ will denote the leader selection function. $F$ can be any function $F: \mathbb{N}  \rightarrow \{ 1, ..., n\}$ for which $\forall i \in \{1, ..., n\}, \# F^{-1}(i) = \infty$. Usually $F(r) = r \mod n$\\

\textbf{Lemma 3}: For any correct process $P_j$, $Q_i.Enqueue(s, m)$ is called by $P_j$ if and only if $m = output(VCBC_i^s, j) \wedge m \ne \bot$

\begin{proof}
    \step{}{If $Q_i.Enqueue(s, m)$ is called by $P_j$ then $m = output(VCBC_i^s, j) \wedge m \ne \bot$}
    \begin{proof}
        \pf $P_j$ is correct (follows protocol) and $Q_i.Enqueue(s, m)$ is only called on line 18 of the broadcast component upon output of $m$ by $VCBC_i^s$ (thus $output(VCBC_i^s, j) \ne \bot$)
    \end{proof}
    
    \step{}{If $m = output(VCBC_i^s, j) \wedge m \ne \bot$ then $Q_i.Enqueue(s, m)$ is called by $P_j$}
    \begin{proof}
        \pf $P_j$ is correct (follows protocol). If $m \ne \bot$, then $VCBC_i^s$ delivered $m$ to $P_j$. Thus, $P_j$ called $Q_i.Enqueue(s, m)$ (lines 16-18 of the broadcast component)
    \end{proof}

    \qedstep{}{
        \begin{proof}
            \pf By 1 and 2        
        \end{proof}
    }
\end{proof}

\textbf{Lemma 4}: If $Q_i.Dequeue(m)$ is called by a correct process $P_j$ at round $r'$, then $\forall r > r', m \in S_j^r$.

\begin{proof}
    \step{}{Once $m$ is added to $S_j$ by $P_j$, it is not removed} 
    \begin{proof}
        \pf By analysis of protocol \reviewdel{specification} \review{pseudo-code} and fact that process $P_j$ follows protocol
    \end{proof}

    \step{}{$Q_i.Dequeue(m)$ is only called on line 20 of broadcast component and line 29 of agreement component}
    \begin{proof}
        \pf By analysis of protocol \reviewdel{specification} \review{pseudo-code} and fact that process $P_j$ follows protocol
    \end{proof}

    \step{}{$m \in S_j^{r'+1}$}
    \begin{proof}
    \step{}{\case{$Q_i.Dequeue(m)$ is called on line 20 of broadcast component}}
    \begin{proof}
        \pf When line 20 of broadcast component is reached at round $r'$, $m \in S_j^{r'}$, which implies, by 1 that $m \in S_j^{r'+1}$
    \end{proof}

    \step{}{\case{$Q_i.Dequeue(m)$ is called on line 29 of agreement component}}
    \begin{proof}
        \pf When line 29 of agreement component is reached at round $r'$, $m$ is added after to $S_j$ if it is not already there,  which implies, that $m \in S_j^{r'+1}$
    \end{proof}

    \qedstep{}{
        \begin{proof}
            \pf By 3.1, 3.2 and 2
        \end{proof}    
    }
    \end{proof}
    \qedstep{}{
        \begin{proof}
            \pf By 1 and 3
        \end{proof}
    }
\end{proof}

\textbf{Lemma 5}: For any correct process $P_j$, if $m \in S_j^r$, then at \review{the beginning of} round $r$, $\forall i, m \notin Q_i$ in process $j$.

\begin{proof}
    \step{}{When $m$ is added to $S_j$, it's removed from every queue in process $P_j$}
    \begin{proof}
        \pf By lines 26-30 of agreement component and fact that $P_j$ is correct
    \end{proof}

    \step{}{If $m \in S_j^r$, then at round $r$, it's not added to any queue}
    \begin{proof}
        \pf By lines 18-20 of broadcast component, atomicity assumption on broadcast actions and fact that process $P_j$ is correct.
    \end{proof}
    
    \qedstep{}{
        \begin{proof}
            \pf By 1 and 2
        \end{proof}
    }
\end{proof}

\textbf{Lemma 6}: For a correct process $P_j$, the following two conditions are equivalent:

\begin{enumerate}[label=(\roman*)]
    \itemsep0em 
    \item $m = value_j(i, s) \wedge m \ne \bot \wedge s = choice_j(i, r)$
    \item $s$ is $Q_i$'s head and $Q_i.Peek() \ne \bot$ at the start of round $r$
\end{enumerate}

\begin{proof}
    \step{}{If (i) then (ii)}
    \begin{proof}
        \step{}{
            \assume{\begin{pfenum}
                \item $m = value_j(i, s)$
                \item $m \ne \bot$
                \item $s = choice_j(i, r)$
            \end{pfenum}}
            \prove{$s$ is $Q_i$'s head and $Q.Peek() \ne \bot$ at the start of round $r$}
        }

        \step{}{$s \ne \bot$}
        \begin{proof}
            \pf By assumptions 1.1.1, 1.1.2 and 1.1.3, and Definition 3
        \end{proof}
    
        \step{}{$\forall s' \le s, \exists m': Q_i.Enqueue(s', m')$ was called by $P_j$ (before the beginning of round $r$)}
        \begin{proof}
            \pf By Lemma 3, Definition 2 (properties 1 and 3), and step 1.2
        \end{proof}
        
        \step{}{$\forall s' < s, \exists m': Q_i.Enqueue(s', m')$ was called $P_j$ and later $Q_i.Dequeue(m')$ was called by $P_j$ (before the beginning of round $r$)}
        \begin{proof}
            \step{}{If $m \in S_j^r$ and $Q_i.Enqueue(s, m)$ is called by $P_j$, $Q_i.Dequeue(m)$ must have been also called \reviewdel{(}before the beginning of round $r$}
            \begin{proof}
                \step{}{
                    \assume{\begin{pfenum}
                        \item $m \in S_j^r$
                        \item $Q_i.Enqueue(s, m)$ was called by $P_j$
                        \item $Q_i.Dequeue(m)$ was not called by $P_j$ \review{before the beginning of round $r$}
                    \end{pfenum}}
                    \prove{False}
                }

                \step{}{$m \notin Q_i$ \review{at beginning of round $r$}}
                \begin{proof}
                    \pf By assumption 1.4.1.1.1 and Lemma 5
                \end{proof}
                \step{}{$m \in Q_i$ \review{at beginning of round $r$}}
                \begin{proof}
                    \pf By assumptions 1.4.1.1.2 and 1.4.1.1.3, and priority queue construction \review{(if $m$ is enqueued, and not dequeued, then $m \in Q_i$)}
                \end{proof}
                \qedstep{}{
                    \begin{proof}
                        \pf By contradiction between 1.4.1.2 and 1.4.1.3
                    \end{proof}
                }
                
            \end{proof}

            \qedstep{}{
                \begin{proof}
                    \pf By steps 1.4.1 and 1.3
                \end{proof}
            }
        \end{proof}
        
        \step{}{$Q_i.Dequeue(m)$ has not been called by $P_j$ after calling $Q_i.Enqueue(s, m)$ (before the beginning of round $r$)}
        \begin{proof}
            \step{}{$m \notin S_j^r$}
            \begin{proof}
                \pf By assumptions 1.1.1 and 1.1.3, step 1.2, and Definitions 2 (property 4) and 3
            \end{proof}

            \qedstep{}{
                \begin{proof}
                    By 1.5.1 and Lemma 4
                \end{proof}
            }
        \end{proof}
        
        \qedstep{}{
        \begin{proof}
            \pf \review{By 1.3 and 1.4, the values in slots $s' < s$ were enqueue and dequeued. By 1.5 the value in $s$ has not yet been dequeued, so $s$ is the head of the queue. By 1.3, some value was enqueued in $s$, so $Q_i.Peek() \ne \bot$.} \reviewdel{ By 1.3, 1.4, 1.5, and priority queue construction }
        \end{proof}
        }
    \end{proof}
    
    \step{}{If (ii) then (i)}
    \begin{proof}
        \step{}{
            \assume{\begin{pfenum}
                \item $s$ is $Q_i$'s head (at the start of round $r$)
                \item $Q_i.Peek() \ne \bot$ (at the start of round $r$)
            \end{pfenum}}
            \prove{$m = value_j(i, s) \wedge m \ne \bot \wedge s = choice_j(i, r)$}
        }
        \step{}{$\forall s' < s, \exists m': Q_i.Enqueue(s', m')$ was called by $P_j$ and after that $Q_i.Dequeue(m')$ was called by $P_j$ (before the beginning of round $r$)}
        \begin{proof}
            \pf \review{The head is by definition the first slot whose value was not dequeued. Therefore, for all slots $s' < s$, the value in $s'$ was dequeued. Since the value of a slot is only dequeued if it was previously enqueued, then for all slots $s' < s$, some value was enqueued to $s'$.} \reviewdel {By assumption 2.1.1 and construction of the priority queue }
        \end{proof}
        
        \step{}{$\forall s' < s, output(VCBS_i^s, j) \ne \bot$ (at the start of round $r$)}
        \begin{proof}
            \pf By Lemma 3 and 2.2
        \end{proof}
        
        \step{}{$\forall s' < s, output(VCBC_i^{s'}, j) \in S_j^r$}
        \begin{proof}
            \pf By Lemma 4 and 2.2
        \end{proof}
        
        \step{}{$Q_i.Enqueue(s, m)$ was called by $P_j$ (before the beginning of round $r$)}
        \begin{proof}
            \pf By assumptions 2.1.1 and 2.1.2 and construction of the priority queue \review{($Q_i.Peek()$ returns the value of the head, and for the head to have a value, $Q_i.Enqueue(s, m)$ must have been called)}
        \end{proof}
        
        \step{}{$m = output(VCBC_i^s, j) \land m \ne \bot$ (at the start of round $r$)}
        \begin{proof}
            \pf By step 2.5 and Lemma 3 
        \end{proof}
        
        \step{}{$m \in Q_i$ (at the start of round $r$)}
        \begin{proof}
            \pf \review{By 2.1.1 and 2.1.2, the value of $s$ has not yet been dequeued (otherwise $s$ would not be the head of the queue). By 2.5, a value has been enqueued to $s$. For these reasons, the value in $s$ is still in $Q_i$.} \reviewdel{By assumptions 2.1.1 and 2.1.2, step 2.5, and construction of the priority queue.}
        \end{proof}
        
        \step{}{$output(VCBC_i^s, j) \notin S_j^r$}
        \begin{proof}
            \pf By 2.7 and Lemma 5 \reviewdel{(modus tollens)}
        \end{proof}
        
        \step{}{$s = choice_j(i, r) \land s \ne \bot$}
        \begin{proof}
            \pf By 2.3, 2.4, 2.6, 2.8 and Definition 2
        \end{proof}
        
        \qedstep{}{
            \begin{proof}
                \pf By 2.9, 2.6 and Definition 3
            \end{proof}
        }
    
    \end{proof}
    
    \qedstep{}{
        \begin{proof}
            \pf By 1 and 2
        \end{proof}
    }
\end{proof}

\textbf{Lemma 7}: For any correct process $P_j$, the following two conditions are equivalent:

\begin{enumerate}[label=(\roman*)]
    \itemsep0em 
    \item $P_j$ delivers $m$ at round $r$
    \item $m = value_j(F(r), r) \wedge m \ne \bot \wedge output(ABA_r, j) = 1$
\end{enumerate}

(for convenience, let $i = F(r)$)

\begin{proof}
    \step{}{If (i) then (ii)}
    \begin{proof}
        \step{}{
            \assume{\begin{pfenum}
                \item $P_j$ delivers $m$ at round $r$
            \end{pfenum}}
            \prove{$m = value_j(i, r) \wedge m \ne \bot \wedge output(ABA_r, j) = 1$}
        }
        \step{}{$output(ABA_r, j) = 1$}
        \begin{proof}
            \pf $P_j$ correct, and delivering a message (assumption 1.1.1) required calling \texttt{AC-Deliver}, which is only done in line 15 of agreement component, under the condition that $ABA_r$ delivers 1 (line 11)
        \end{proof}
        
        \step{}{$Q_i.Peek() = m \wedge m \ne \bot$}
        \begin{proof}
            \pf $P_j$ correct, and delivering a message (assumption 1.1.1) required calling \texttt{AC-Deliver}, which is only done in line 15 of agreement component with $Q_i.Peek()$, under the condition that $Q_i.Peek() \ne \bot$ (line 14).
        \end{proof}
        
        \step{}{$m = value_j(i, r)$}
        \begin{proof}
            \pf By \reviewdel{construction of the priority queues, } step 1.3, and Lemma 6 \review{(with $m = Q_i.Peek()$)}
        \end{proof}
        
        \qedstep{}{
            \begin{proof}
                \pf By 1.1, 1.2, 1.3 and 1.4
            \end{proof}
        }
    \end{proof}
    
    \step{}{If (ii) then (i)}
    \begin{proof}
        \step{}{
            \assume{\begin{pfenum}
                \item $m = value_j(i, r)$ 
                \item $m \ne \bot$
                \item $output(ABA_r) = 1$
            \end{pfenum}}
            \prove{$P_j$ delivered $m$ at round $r$}
        }
        \step{}{$Q_i.Peek() = m \wedge m \ne \bot$}
        \begin{proof}
            \pf By assumptions 2.1.1 and 2.1.2, and Lemma 6
        \end{proof}
        
        \qedstep{}{
            \begin{proof}
                \pf By steps 2.2 and 2.1.3, and lines 10-15 of agreement component
            \end{proof}
        }
    \end{proof}
    
    \qedstep{}{
        \begin{proof}
            \pf By 1 and 2
        \end{proof} 
    }
\end{proof}

\textbf{Lemma 8}: A correct process $P_j$ \reviewdel{votes} \review{inputs} 1 for $ABA_r$ if and only if $choice_j(F(r), r) \ne \bot$ at the time of voting

(for convenience, let $i = F(r)$)

\begin{proof}
    \step{}{If a correct process $P_j$ \reviewdel{votes} \review{inputs} 1 for $ABA_r$ then $choice_j(i, r) \ne \bot$ at the time of \reviewdel{voting} \review{inputting}}
    \begin{proof}
        \step{}{$Q_i.Peek() \ne \bot$ (at the start of round $r$)}
        \begin{proof}
            \pf Lines 6-9 of agreement component
        \end{proof}

        \qedstep{}{
            \begin{proof}
                \pf By 1.1 and Lemma 6
            \end{proof}
        }
    \end{proof}
    
    \step{}{If $choice_j(i, r) \ne \bot$ at the time of \reviewdel{voting} \review{inputting} then a correct process $P_j$ \reviewdel{votes} \review{inputs} 1 for $ABA_r$ }
    \begin{proof}
        \review{
        \step{}{\assume{\begin{pfenum}
            \item $P_j$ a correct process
            \item $choice_j(i, r) \ne \bot$ at the time of inputting for $ABA_r$
        \end{pfenum}}
        \prove{$P_j$ inputs 1 for $ABA_r$}}
        }
        \step{}{$Q_i.Peek() \ne \bot$}
        \begin{proof}
            \pf By Lemma 6 and \review{2.1.1} \reviewdel{$choice_j(i, r) \ne \bot$}
        \end{proof}

        \qedstep{}{
            \begin{proof}
                \pf By lines 8 and 9 of agreement component and 2.\reviewdel{1}\review{2}
            \end{proof}
        }
        
    \end{proof}
    
    \qedstep{}{
        \begin{proof}
            \pf By 1 and 2    
        \end{proof}
    }
    
\end{proof}

\textbf{Lemma 9}: Let $P_j$ and $P_k$ be correct processes. If $P_j$ delivers at round $r'$, and $P_k$ reaches round $r$ and $r > r'$, then $P_k$ delivered at round $r'$.

\begin{proof}
    \step{}{
        \assume{\begin{pfenum}
            \item $P_j$ correct
            \item $P_k$ correct
            \item $P_j$ delivers at round $r'$        
            \item $P_k$ reaches round $r$
            \item $r > r'$
        \end{pfenum}}
        \prove{$P_k$ delivers at round $r'$}
    }
    \step{}{$output(ABA_{r'}, j) = 1$}
    \begin{proof}
        \review{
            \step{}{\texttt{AC-Deliver} is only called if $output(ABA_r, j) = 1$}
            \pf{By lines 11-15 of agreement component}
        }

        \qedstep{}{
            \pf \reviewdel{\texttt{AC-Deliver} is only called if $output(ABA_r, j) = 1$}\review{2.1}, and assumptions 1.1 and 1.3.
        }
    \end{proof}

    \step{}{$output(ABA_{r'}, k) \ne \bot$}
    \begin{proof}
        \step{}{$P_k$ moved to round $r'+1$}
        \begin{proof}
            \pf $P_k$ is correct (1.2), is at round $r$ ($r > r'$, by 1.5) and round numbers increase one a time (line 16 of agreement component)
        \end{proof}
        
        \qedstep{}{
            \begin{proof}
                \pf By line 10 of agreement component, 1.2 and 3.1
            \end{proof}
        }
    \end{proof}
    
    \step{}{$output(ABA_{r'}, k) = 1$}
    \begin{proof}
        \pf By 1.1, 1.2, 2, 3 and ABA's agreement property
    \end{proof}
    
    \step{}{If round $r'$ terminates for correct process $P_k$ and $output(ABA_{r'}, k) = 1$, then $P_k$ delivers some message $m$}
    \begin{proof}
        \pf Lines 11-15 of agreement component
    \end{proof}
    
    \qedstep{}{
        \begin{proof}
            \pf By 5, 4 and assumptions 1.2, 1.4 and 1.5
        \end{proof}
    }
\end{proof}

\textbf{Lemma 10}: Let $P_j$ be a correct process that is at round $r$. Eventually $P_j$ finishes round $r$ and moves to round $r+1$.

\begin{proof}
    \step{}{$ABA_r$ eventually decides some value $b \in \{0, 1\}$}
    \begin{proof}
        \pf By ABA's termination property
    \end{proof}

    \qedstep{}{}
    \begin{proof}
        \step{}{\case{$b = 0$}}
        \begin{proof}
            \pf The round terminates immediately after $ABA_{r}$ decides (lines 10-16 of agreement component)
        \end{proof}
         
        \step{}{\case{$b = 1$}}
        \begin{proof}
            \step{}{\case{$Q.Peek() \ne \bot$}}
            \begin{proof}
                \pf The round terminates immediately (lines 12-16,25-31 of agreement component)
            \end{proof}
            
            \step{}{\case{$Q.Peek() = \bot$ (for convenience, let $Q.head = s$ and $i = F(r)$)}}
            \begin{proof}
                \step{}{At least one correct process $P_k$ \reviewdel{voted} \review{inputted} 1 for $ABA_r$}
                \begin{proof}
                    \pf By ABA's validity property, if all correct processes had \reviewdel{voted} \review{inputted} 0, $output(ABA_r, j) = 0$, which is not the case, by assumption.
                \end{proof}

                \step{}{$P_k$ will reply with valid \texttt{FILLER} request to $P_j$}
                \begin{proof}
                    \step{}{For $P_k$, $Q_i.head \ge s$}
                    \begin{proof}
                        \pf At the time of voting of round $r$, $P_k$'s $Q_i.head$ was $s$ and the head's priority does not decrease by construction.
                    \end{proof}

                    \step{}{$P_k$ has a valid proof for $VCBC_s^i$}
                    \begin{proof}
                        \pf By fact that $P_k$ \reviewdel{voted} \review{inputted} 1, Lemma 8 and VCBC's verifiability property
                    \end{proof}

                    \qedstep{}{
                        \begin{proof}
                            \pf By 2.2.2.2.1, 2.2.2.2.2 and lines 17-21 of agreement component
                        \end{proof}
                    }
                \end{proof}

                \step{}{Eventually, $output(VCBC_s^i, j) \ne \bot$}
                \begin{proof}
                    \pf By 2.2.2.2 and lines 22-24 of agreement component
                \end{proof}
                
                \step{}{$Q.Peek() \ne \bot$ eventually for $P_j$}
                \begin{proof}
                    \pf By 2.2.2.3
                \end{proof}
                
            \end{proof}

            \qedstep{}{
                \begin{proof}
                    \pf By 2.2.2.4 and lines 14-16,25-31 of agreement component
                \end{proof}
            }
        \end{proof}
    
        \qedstep{}{}
    \end{proof}
\end{proof}

\textbf{Theorem 1}: Let $P_j$ and $P_k$ be correct processes and $holds(r) = True$.
$P_j$ delivers $m$ at $r$ $\implies$ $P_k$ delivers $m$ at $r$

\begin{proof}
    \step{}{\assume{\begin{pfenum}
        \item $P_j$ correct
        \item $P_k$ correct
        \item $holds(r) = True$
        \item $P_j$ delivers $m$ at $r$ 
    \end{pfenum}}
    \prove{$P_k$ delivers $m$ at $r$}
    }

    \step{}{$P_k$ eventually reaches round $r+1$}
    \begin{proof}
        \pf By 1.2 and Lemma 10
    \end{proof}

    \step{}{$P_k$ delivers some $m'$ at round $r$}
    \begin{proof}
        \pf By Lemma 9, assumptions 1.1, 1.2 and 1.4, and step 2
    \end{proof}

    \step{}{$m' = value_k(i, r) \ne \bot$}
    \begin{proof}
        \pf By Lemma 7, assumption 1.2 and step 3
    \end{proof}

    \step{}{$m = value_j(i, r) \ne \bot$}
    \begin{proof}
        \pf By Lemma 7 and assumptions 1.2 and 1.4
    \end{proof}

    \qedstep{}{
        \begin{proof}
            \pf By Lemma 2, assumptions 1.1, 1.2 and 1.3, step 4, and Definition 3
        \end{proof}
    }
\end{proof}

\textbf{Lemma 11}: $holds(r) \implies holds(r+1)$

\begin{proof}
    \step{}{
        \assume{\begin{pfenum}
            \item $P_j$ correct
            \item $P_k$ correct
            \item $S_j^r = S_k^r$
        \end{pfenum}}
        \prove{$S_j^{r+1} = S_k^{r+1}$}
    }
    \qedstep{}{
    }
    \begin{proof}
        \step{}{
            \case{Some $m$ is delivered by $P_j$ at $r$}
            \begin{proof}
                \step{}{$S_j^{r+1} = S_j^r \cup \{m\} $}
                \begin{proof}
                    \pf By 2.1
                \end{proof}
                
                \step{}{$P_k$ delivers $m$ at $r$}
                \begin{proof}
                    \pf By Theorem 1
                \end{proof}

                \step{}{$S_k^{r+1} = S_k^r \cup \{m\}$}
                \begin{proof}
                    \pf By 2.1.2
                \end{proof}
                
                \qedstep{}{
                    \begin{proof}
                        \pf $S_j^{r+1} = S_j^r \cup \{m\} = S_k^r \cup \{m\} = S_k^{r+1}$
                    \end{proof}                
                }
            \end{proof}
        }
        
        \step{}{
            \case{$P_j$ delivers no message in round $r$}
            \begin{proof}
                \step{}{$S_j^{r+1} = S_j^r$}
                \begin{proof}
                    \pf Because no message was delivered by $P_j$
                \end{proof}
                
                \step{}{$P_k$ delivers no message in round $r$}
                \begin{proof}
                    \step{}{
                        \assume{\begin{pfenum}
                            \item $P_k$ delivers message $m$ in round $r$
                        \end{pfenum}}
                        \prove{$False$}
                    }
                    \qedstep{}{
                        \begin{proof}
                            \step{}{$P_j$ delivers $m$ in round $r$}
                            \begin{proof}
                                \pf By Theorem 1 and assumptions 1.1, 1.2, 1.3 (consensus holds) and 2.2.2.1.1
                            \end{proof}
    
                            \qedstep{}{
                                \begin{proof}
                                    \pf Contradiction between 2.2.2.2.1 and 2.2
                                \end{proof}
                            }
                        \end{proof}
                    }
                \end{proof}
                
                \step{}{$S_k^{r+1} = S_k^r$}
                \begin{proof}
                    \pf By 2.2.2
                \end{proof}
                
                \qedstep{}{
                    \begin{proof}
                        \pf $S_j^{r+1} = S_j^r = S_k^r = S_k^{r+1}$
                    \end{proof}                
                }
            \end{proof}
        }
        
        \qedstep{}{
            \begin{proof}
                \pf By 2.1 and 2.2
            \end{proof}
        }
    \end{proof}
\end{proof}

\textbf{Theorem 2}: $\forall r, holds(r) = True$

\begin{proof}
    \step{}{$\forall P_j$ correct, $S_0^j = \{\}$} 
    \begin{proof}
        \pf By line 5 of protocol initialization
    \end{proof}

    \step{}{$holds(0) = True$}
    \begin{proof}
        \pf By step 1 and Definition 1
    \end{proof}

    \step{}{$holds(r) \implies holds(r+1)$}
    \begin{proof}
        \pf By Lemma 11
    \end{proof}

    \qedstep{}{
        \begin{proof}
            \pf By induction on $r$, with base case proved by 2 and induction step by 3 
        \end{proof}
    }
\end{proof}

\textbf{Corollary 1}: Given $P_j$ and $P_k$ correct processes, $P_j$ delivers $m$ at $r$ $\implies$ $P_k$ delivers $m$ at $r$

\begin{proof}
    \step{}{$holds(r) = True$}
    \begin{proof}
        \pf By Theorem 2
    \end{proof}

    \qedstep{}{By step 1 and Theorem 1}
\end{proof}

\textbf{Corollary 2}: Given $P_j$ and $P_k$ two correct process. $P_j$ delivers $m$ $\implies$ $P_k$ delivers $m$

\begin{proof}
    \step{}{$P_j$ delivered $m$ at some round $r$}
    \begin{proof}
        \pf Process is always at some round
    \end{proof}

    \qedstep{}{
        \begin{proof}
            \pf By step 1 and Corollary 1
        \end{proof}
    }
\end{proof}

\textbf{Theorem 3}: For any message $m$, a correct process $P_j$ delivers $m$ at most once.

\begin{proof}
    \step{}{$m$ delivered by $P_j \implies m \notin S_j$} 
    \begin{proof}
        \pf By lines 29 and 31 of agreement component (delivery only occurs in line 31).
    \end{proof}

    \step{}{When $m$ is delivered by $P_j$, $m$ is added to $S_j$}
    \begin{proof}
        \pf By lines 29-31 of agreement component (delivery only occurs in line 31).
    \end{proof}

    \step{}{Once $m$ is added to $S_j$, $m$ is never removed}
    \begin{proof}
        \pf Elements are never taken from $S_j$.
        \mfnote{I think that it may be better so use lemma 5 as proof here. What do you think?}
        \dsanote{Don't think so}        
    \end{proof}

    \qedstep{}{
        \begin{proof}
            \pf By steps 1-3.
        \end{proof}
    }
\end{proof}

\textbf{Definition 4}: Given a node $P_i$ and a round $r$, $next(i, r)$ denotes the smallest round $r'$
such that $F(r') = i \wedge r' \ge r$

$next(i, r)$ exists for all $i$ and $r$, because $\# F^{-1}(i) = \infty$ (by assumption on $F$).

\textbf{Lemma 12}: For any $m$ and $r$, $\forall P_j $ correct, $value_j(i, r) = m~\wedge~m \ne \bot \implies m$ is delivered at some $r'$ with $r \le r' \le next(i, r)$.

\begin{proof}
    \step{}{
        \assume{\begin{pfenum}
            \item $m \ne \bot$   
            \item $\forall P_j$ correct, $value_j(i, r) = m$
        \end{pfenum}}
        \prove{$m$ is delivered at some round $r'$ with $r \le r' \le next(i, r)$} 
    }

    \qedstep{}{}
    \begin{proof}
        \step{}{
            \case{$m$ is delivered at $r'$ with $r \le r' < next(i, r)$}
        }
        \begin{proof}
            \pf Trivially, $m$ is delivered at some round $r'$ with $r \le r' \le next(i, r)$ (because $r \le r' < next(i, r) \implies r \le r' \le next(i, r)$)
        \end{proof}
        
        \step{}{
            \case{$m$ is not delivered before $next(i, r)$}
        }

        \begin{proof}
            \step{}{
                \sassume{\begin{pfenum}
                    \item $m \ne \bot$
                    \item $P_j$ is at round $next(i, r)$
                    \item $m$ was not delivered by $P_j$ as of round $next(i,r)-1$
                    \item $\forall P_j$ correct, $value_j(i, r) = m$
                \end{pfenum}
                }
                \prove{$P_j$ delivers $m$ at round $next(i, r)$}}

            \step{}{$\forall P_j$ correct, $value_j(i, next(i, r)) = m$. (For convenience, let s = $choice_j(r, i)$)}

            \begin{proof}
                \step{}{$\forall s' < s, output(VCBC_i^{s'}, j) \ne \bot$}
                \begin{proof}
                    \pf By assumptions 2.2.1.4 and 2.2.1.1, and Definitions 2 (property 1) and 3.
                \end{proof}                    
                
                \step{}{$\forall s' < s, output(VCBC_i^{s'}, j) \in S_{next(i, r)}^j$}
                \begin{proof}
                    \pf By assumptions 2.2.1.4 and 2.2.1.1, Definitions 2 (property 2) and 3, and the fact that nothing is removed from $S_j$.
                \end{proof}

                \step{}{$output(VCBC_i^s, j) \ne \bot$}
                \begin{proof}
                    \pf By assumptions 2.2.1.1 and 2.2.1.4, and Definitions 2 (property 3) and 3.
                \end{proof}

                \step{}{$output(VCBC_i^s, j) \notin S_{next(i, r)}^j$}
                \begin{proof}
                    \pf By assumptions 2.2.1.1 - 2.2.1.4 and Definitions 2 (property 4) and 3.
                \end{proof}

                \qedstep{}{
                    \begin{proof}
                        \pf By 2.2.2.1 - 2.2.2.4 and Definition 2
                    \end{proof} 
                }
            \end{proof}

            \step{}{$\forall P_j$ correct, $P_j$ \reviewdel{votes} \review{inputs} 1 for $ABA_{next(i, r)}$}
            \begin{proof}
                \pf By step 2.2.2 and Lemma 8
            \end{proof}
            
            \step{}{$\forall P_j$ correct, $output(ABA_{next(i, r)}, j) = 1$}
            \begin{proof}
                \pf By step 2.2.3 and ABA's validity and termination properties.
            \end{proof}
            
            \step{}{$\forall P_j$ correct, $P_j$ delivers $m$ at $next(i, r)$}
            \begin{proof}
                \pf By step 2.2.4 and Lemma 7
            \end{proof}

            \qedstep{}{
                \begin{proof}
                    \pf By step 2.2.1 and 2.2.5
                \end{proof} 
            }
        \end{proof}

        \qedstep{}{
            \begin{proof}
                \pf By 2.1 and 2.2    
            \end{proof}
        }
    \end{proof}
\end{proof}

\textbf{Theorem 4}: If a correct process $P_j$ receives a request $m$ from a client, then all correct processes eventually deliver $m$.

\begin{proof}
    \qedstep{}{}
    \begin{proof}
        \step{}{\case{$m \in S_j$ (line 10 of broadcast component)}}
        \begin{proof}
            \pf If $m \in S_j$, then $P_j$ delivered $m$. Then all processes delivered by Corollary 2
        \end{proof}
        
        \step{}{\case{$m \notin S_j$ (line 10 of broadcast component)}}
        \begin{proof}
            \step{}{$m$ is inputted into a VCBC instance - $VCBC_s^j$ for some slot $s$}
            \begin{proof}
                \pf By lines 9-13 of broadcast component protocol
            \end{proof}

            \step{}{$\forall s' \le s, \exists m': \forall P_k$ correct, $Q_j.Enqueue(s', m')$ is called by $P_k$} 
            \begin{proof}
                \step{}{
                    \sassume{\begin{pfenum}
                        \item $s' \le s$ 
                        \item $s'$ is smallest slot for which $Q_j.Enqueue(s', m)$ was not called by some process $P_k$
                    \end{pfenum}}
                    \prove{False}
                }

                \step{}{$VCBC_{s}^j$ started $\implies$ $VCBC_{s-1}^j$ started $\vee$ $s = 0$}
                \begin{proof}
                    \pf Lines 13-15 of broadcast component
                \end{proof}

                \step{}{$VCBC_{s'}^j$ was started by $P_j$}
                \begin{proof}
                    \pf By 1.2.1 and 1.2.2.2
                \end{proof}
                
                \step{}{By some round $r$, $VCBC_{s'}^j$ delivers to all correct processes $P_k$}
                \begin{proof}
                    \pf By VCBC's validity property and fact that $P_j$ is correct
                \end{proof}

                \step{}{$\forall P_k$ correct, $Q_j.Enqueue(s', m')$ is called}
                \begin{proof}
                    \pf By 1.2.2.4 and Lemma 3
                \end{proof}

                \qedstep{}{
                    \begin{proof}
                        \pf By contradiction reached in 1.2.2.5
                    \end{proof}
                }
            \end{proof}

            \step{}{$\forall s' \le s$, the value input to $VCBC_{s'}^j$ is eventually AC-delivered\footnote{We use AC-deliver in this section of the proof to distinguish between a delivery of a totally-ordered broadcast value from a delivery of the VCBC sub-protocol.} by all correct processes}
            \begin{proof}
                \step{}{
                    \sassume{\begin{pfenum}
                        \item $s' \le s$
                        \item $s'$ is first slot for which the value input to $VCBC_{s'}^j$ is never AC-delivered by some correct process $P_k$
                    \end{pfenum}}
                    \prove{False}
                }

                \step{}{
                    $\forall s'' < s'$, $s''$'s value is AC-delivered by all correct process $P_k$
                    (let $r$ be the round after $s'-1$ is delivered)
                }
                \begin{proof}
                    \pf By assumption 1.2.3.1.2
                \end{proof}                

                \step{}{
                    $VCBC_{s'}^j$ delivers to all $P_k$ by some round $r'$
                }
                \begin{proof}
                    \pf By VCBC's validity property and the fact that $P_j$ is correct
                \end{proof}

                \step{}{$s'$'s value is eventually AC-delivered by all processes}
                \begin{proof}
                    \step{}{\case{$m$ is AC-delivered before $max(r, r')$}}
                    \begin{proof}
                        \pf Then obviously, $s'$'s value is eventually AC-delivered
                    \end{proof}
                    
                    \step{}{\case{$m$ has not been AC-delivered by $max(r, r')$}}
                    \begin{proof}
                        \step{}{At $max(r, r'), \forall P_k$ correct, $choice_k(j, max(r, r')) = s'$ and $value_k(j, max(r, r')) = m'$} 
                        \begin{proof}
                            \pf By 1.2.3.2, 1.2.3.3, 1.2.3.4.2, and Definition 2
                        \end{proof}

                        \step{}{$m$ is delivered at some $r''$ with $max(r, r') < r'' < next(j, max(r, r'))$}
                        \begin{proof}
                            \pf By 1.2.3.4.2.1 and Lemma 12
                        \end{proof}

                        \qedstep{}{
                            \begin{proof}
                                \pf By 1.2.3.4.2.2
                            \end{proof}
                        }
                    \end{proof}
                    
                    \qedstep{}{
                        \begin{proof}
                            \pf By 1.2.3.4.1 and 1.2.3.4.2
                        \end{proof}
                    }
                \end{proof}

                \qedstep{}{
                    \begin{proof}
                        \pf By the contradiction between assumption 1.2.3.1.2 and step 1.2.3.4
                    \end{proof}
                }
            \end{proof}

            \qedstep{}{
                \begin{proof}
                    \pf By 1.2.3
                \end{proof}
            }
        \end{proof}
        
        \qedstep{}{
            \begin{proof}
                \pf By 1.1 and 1.2
            \end{proof}
        }
    \end{proof}
\end{proof}

\textbf{Theorem 5}: If correct processes $P_j$ and $P_k$ deliver $m$ and $m'$, then they do so in the same order.

\begin{proof}
    \step{}{Only one round is executed by a process at a time and processes go through the rounds incrementally, one by one.}
    \begin{proof}
        \pf By construction of the agreement component.
    \end{proof}

    \step{}{If $P_j$ delivered $m$ at $r$, then $P_k$ delivered $m$ at $r$}
    \begin{proof}
        \pf By Corollary 1
    \end{proof}

    \step{}{If $P_j$ delivered $m'$ at $r'$, then $P_k$ delivered $m'$ at $r'$}
    \begin{proof}
        \pf By Corollary 1
    \end{proof}

    \qedstep{}{
        \begin{proof}
            \pf By steps 1-3
        \end{proof}
    }
\end{proof}

\end{document}